\documentclass[aps,floats,amsmath,amssymb,twocolumn,superscriptaddress,preprintnumbers,nofootinbib]{revtex4-1}
\usepackage{latexsym}
\usepackage{amssymb,amsmath}
\usepackage{graphicx}
\usepackage{color}
\usepackage{hyperref}

\newcommand{\be}{\begin{equation}}
\newcommand{\ee}{\end{equation}}

\def\la{\mathrel{\mathpalette\fun <}}
\def\ga{\mathrel{\mathpalette\fun >}}
\def\fun#1#2{\lower3.6pt\vbox{\baselineskip0pt\lineskip.9pt
  \ialign{$\mathsurround=0pt#1\hfil##\hfil$\crcr#2\crcr\sim\crcr}}}

\def\4he{$^4$He}
\newcommand{\epm}{\ensuremath{e^{\pm}\;}}
\def\omegal{$\Omega_{L}$}

\def\omegalh{$\Omega_{L}h^{2}$}
\def\ml{$m_{L}$}
\def\m100{$m_{100}$}
\def\ql{$Q_{L}$}
\def\L{$L$}
\def\Lpm{$L^{\pm}$}
\newcommand{\eeql}[1]{\label{#1}\eeq}

\newcommand{\ra}{\ensuremath{\rightarrow}}

\newcommand{\skipblk}[1]{}                                                      
\def\bqa{\begin{eqnarray}}                                                      
\def\eqa{\end{eqnarray}}                                                        

\newcommand{\etal}{{\em et al., }}

\newcommand{\eg}{{\em e.g., }}                                                  
\newcommand{\ie}{{\em i.e., }}

\newcommand{\sto}{\ensuremath{SU(2) \x U(1)}}                                       
\newcommand{\st}{\ensuremath{SU(2)}}                                                                                       
\newcommand{\sth}{\ensuremath{SU(3)}} 
\newcommand{\stf}{\ensuremath{SU(5)}}                                                                                      
\newcommand{\x}{\ensuremath{\times}}

\newcommand{\beq}{\begin{equation}}                                             
\newcommand{\eeq}{\end{equation}}     
                                          
\newcommand{\oh}{\ensuremath{\frac{1}{2}}}

\begin{document}

\title{Requiem for an FCHAMP?\\ 
(Fractionally CHArged, Massive Particle)}
\author{Paul  Langacker}
\email{pgl@ias.edu}
\affiliation{School of Natural Science, Institute for Advanced Study, Einstein Drive, Princeton, New Jersey 08540, USA}
\affiliation{Department of Physics, Princeton University, Princeton, NJ 08544, USA}
\author{Gary Steigman}
\email{steigman.1@asc.ohio-state.edu}
\affiliation{Departments of Physics and Astronomy, Center for Cosmology and Astro-Particle Physics,\\ 
The Ohio State University, 191 W. Woodruff Ave., Columbus OH 43210-1117, USA}

\begin{abstract}

Fractionally charged massive particles (FCHAMPs) appear in extensions of the standard model, especially those with superstring constructions.  The lightest FCHAMP would be absolutely stable and any of them produced during the early evolution of the Universe would be present today.  The thermal production, annihilation and, survival of an FCHAMP, a lepton \L~with electroweak (i.e., $U(1)_Y$) but no strong interactions, of mass \ml~and charge \ql~(in units of the charge on the electron) are explored.  The FCHAMP relic abundance is determined by the total annihilation cross section which depends on \ml, \ql~and on the available annihilation channels.  Since massive ($m_{L} \ga 1$~GeV) charged particles ($Q_{L} \ga 0.01$) behave like baryons (heavy ions), primordial nucleosynthesis and the cosmic background radiation temperature anisotropies limit the FCHAMP relic density.  Requiring that $\Omega_{L} \la \Omega_{\rm B}/5$ leads to a constraint on the \ql~-- \ml~relation.  Further constraints on \ql~and \ml~are provided by the invisible width of the $Z$ ($Q_L > 0.16$ for $m_{L} \leq M_{Z}/2$) and by accelerator searches for massive, charged particles.  Our key result is to exploit the fact that in the early Universe, after \Lpm~freeze-out but prior to \epm recombination, the negatively charged $L^{-}$ will combine with the more abundant alpha particles and protons to form tightly bound, positively charged states (fractionally charged heavy ions).  The Coulomb barriers between these positively charged $L^{-}\alpha$ and $L^{-}p$ ($L^{-}pp$, $L^{-}\alpha\alpha$, ...) bound states and the free $L^{+}$ suppresses late time FCHAMP annihilation in the interstellar medium (ISM) of the Galaxy and on Earth, limiting significantly the late-time reduction of the FCHAMP abundance compared to its relic value.  The surviving FCHAMP abundance on Earth is orders of magnitude higher than the limits from terrestrial searches for fractionally charged particles, closing the window on FCHAMPs with $Q_{L} \ga 0.01$.  However, as \ql~approaches an integer (\eg $|Q_{L} - n| \la 0.25$) these searches become increasingly insensitive, leaving some potentially unconstrained ``islands" in the \ql~-- \ml~plane which may be explored by searching for FCHAMPs in the cosmic rays.

\end{abstract}

\maketitle

\section{Introduction}
\label{intro}

We consider the cosmological and astrophysical viability of a fractionally charged massive particle, an FCHAMP, the lightest of which is stable on cosmological timescales.  While there is an extensive literature on integer charged CHAMPs and neutraCHAMPs and their neutral counterparts in CHAMP-proton and CHAMP-alpha particle bound states (see, \eg \cite{khlopov,derujula,dimopoulos,gould,davidson}), in very few of these papers were fractionally charged FCHAMPs explored in any detail.  Many of these previous constraints on CHAMPs do not apply to FCHAMPs.  Our study is restricted to a fractionally charged lepton $L$ (an  \Lpm~particle -- antiparticle pair), an $SU(3)$ and $SU(2)$ singlet with weak hypercharge $Y=Q_L$, which couples to the photon and the $Z$.  This choice fixes the total \Lpm~annihilation cross section, determining the relic abundance of \Lpm~pairs in the interstellar medium (ISM) of the Galaxy and on Earth.  Requiring that the relic abundance (mass fraction) be bounded limits a combination of the mass, $m_{L}$, and charge, $Q_{L}$, providing a lower bound to \ql~as a function of \ml~\cite{wolfram,steigman1979,goldberg}.  During the post-annihilation freeze out evolution of the Universe, prior to recombination, the negatively charged $L^{-}$ will combine with alpha partcles and protons to form tightly bound, positively charged, heavy ions such as $L\alpha$, $Lp$, $Lpp$, $L\alpha\alpha$.  The net positive charges of these exotic heavy ions suppresses late time annihilation of the bound $L^{-}$ with the free $L^{+}$.  This suppression leads to a terrestrial ratio of FCHAMPs to baryons which is so high that terrestrial searches for fractionally charged particles~\cite{Perl:2009zz} challenge their very existence.  In \S\ref{motivation} the particle physics motivations for the existence of FCHAMPs are outlined along with general comments on existing experimental limits.  In \S\ref{relic} the relic FCHAMP abundance is calculated and limits are set to $m_{L}$ and $Q_{L}$.  Constraints on \ql~and \ml~from collider searches and from the invisible width of the $Z$ are analyzed in \S\ref{acc}, and those from terrestrial searches for fractionally charged particles in \S\ref{earth}.  The post-annihilation freeze out capture of negatively charged FCHAMPs by protons and alpha particles is described and the ratio of $Lp$ and $L\alpha$ to baryons in the present Universe is estimated in \S\ref{capture}.  In \S\ref{discussion} we compare our results to the bounds from the terrestrial searches to decide if they exclude FCHAMPs -- or not.  Our conclusions are summarized in \S\ref{conclusions}. 

\section{Motivation}
\label{motivation}

All known color-singlet particles have electric charges that are integer multiples of $e$, the charge of the positron. However, it is straightforward to construct theories with fractional or even irrational  charges. Assuming the standard model \sto\ gauge group, the weak hypercharge $Y$ assignments are arbitrary\footnote{Appropriate combinations of non-canonical fermion charge assignments are needed for the cancellation of anomalies.}, as are the associated values of electric charge $Q=T_L^3+Y$, where $T_L^3$ is the third generator of weak isospin.  Even ``ordinary'' values of $Y$ can lead to fractional charges for the ``wrong''  \sth\ or \st\ assignments, such as an \sth\ and \st -singlet state with $Y=\oh$. The hypercharge assignments are no longer arbitrary when the standard model is embedded in a simple grand unified (GUT) group, such as \stf~\cite{Georgi:1974sy}, but  more complicated GUT embeddings allow for fractional charges \cite{GellMann:1976pg}.

Fractionally charged states are extremely common in superstring constructions~\cite{Wen:1985qj,Athanasiu:1988uj}\footnote{This is apparent from the Schellekens theorem~\cite{Schellekens:1989qb}, which essentially states that any weakly-coupled heterotic string theory with realistic gauge couplings must involve either fractional charges, an unbroken \stf\ gauge symmetry, or higher Ka\v c-Moody embeddings of \sth\ or \st. Examples with fractional charges include both heterotic~\cite{Antoniadis:1989zy,Faraggi:1990af,Chaudhuri:1995ve,Cleaver:1998gc} and intersecting D-brane~\cite{Cvetic:2001nr,Cvetic:2002qa,Cvetic:2011iq} constructions. For general discussions, see~\cite{Dienes:1995sq,Lykken:1996kc}. Recent constructions which avoid fractional charges are described in~\cite{Christodoulides:2011zs}.}.  Such fractionally charged particles could acquire very large (string-scale) masses, or could be confined into integer-charged bound states by strongly-coupled hidden sector  forces~\cite{Antoniadis:1989zy}\footnote{Such bound state {\em cryptons}  have been postulated as  sources of ultra high energy cosmic rays~\cite{Ellis:2004cj} or as  dark matter candidates~\cite{Benakli:1998ut,Chang:1996vw,Coriano:2001mg}.}.  However, they could also occur in the massless sector of the theory, e.g., acquiring masses in the TeV range from effects related to electroweak physics.
 
Another motivation is that kinetic mixing between two $U(1)$ gauge bosons~\cite{Holdom:1985ag} can under some circumstances induce small (e.g., millicharged) and generally irrational electric charges for hidden sector particles\footnote{See, e.g, \cite{Holdom:1985ag,Foot:2004pa,Kors:2004dx,Feldman:2007wj}.  Observational constraints are summarized in~\cite{Prinz:1998ua,Davidson:2000hf,haibo}.}. In this paper, however, we will only consider larger charges, $Q_L \ga 0.01$.
 
There have been many searches for fractionally charged particles in fixed target and collider experiments, quark searches (e.g., Millikan-type experiments), cosmic rays, and bulk matter searches, as described in a recent review~\cite{Perl:2009zz}\footnote{Earlier reviews include~\cite{Marinelli:1982dg,smith,Perl:2004qc}.  More general reviews of massive stable particles include~\cite{Perl:2001xi,Fairbairn:2006gg}.}. The lightest fractionally charged particle would have to be absolutely stable (or at least stable on cosmological time scales given the stringent limits on charge non-conservation), and would therefore contribute to the cosmological mass density unless it were so heavy that it was not produced
subsequent to inflation. 
 
Masses smaller than $M_Z/2$ are excluded by the invisible $Z$ width unless the charge is small (e.g., $Q_{L} \la 0.16$; see \S\ref{acc}).  There have been many searches for heavy long-lived particles at accelerators and colliders, but these are difficult to summarize  because the analyses are often presented for specific assumptions concerning the charges and other particle properties. For example,  some limits only apply to fractionally charged particles that carry color, while others are presented for specific models, such as scalar leptons in supersymmetry. Furthermore, some searches are insensitive to particles with charges below a critical value such as $2/3$ because they are not sufficiently ionizing.  Searches for fractionally charged particles produced in \epm annihilation have excluded charges $\ga 2/3$ for masses below $\sim 102$~GeV~\cite{Perl:2009zz,abbiendi}. There are more stringent limits from the Tevatron~\cite{Acosta:2002ju,Abazov:2008qu,Aaltonen:2009kea} and LHC~\cite{Khachatryan:2011ts,Collaboration:2011hz} for heavy quarks or for long-lived integer charged sleptons or charginos.  In particular, a D0 lower limit of 206 GeV on the mass of long-lived charged gauginos~\cite{Abazov:2008qu} will be translated into a limit  on fractionally charged particle masses  in \S\ref{acc}.

For definiteness we consider here the simplest example, that of a new, fractionally charged, massive spin-$1/2$ particle\footnote{The lightest fractionally charged state could also be a scalar, especially in supersymmetric theories. The annihilation of charged scalars into fermion pairs is $P$-wave suppressed, implying even more stringent cosmological and bulk matter  bounds.}, an FCHAMP ``$L$", that is an \sth\ and \st-singlet with weak hypercharge (and therefore electric charge) $Y$, which therefore couples to both the photon and the $Z$. In calculating the FCHAMP relic abundance the total \Lpm~annihilation cross section enters.  In our estimate we consider the dominant annihilation channels $L^+ L^- \ra f \bar f$, where $f$ is a standard model fermion (including the $t$ for $m_L > m_t$), as well as the diboson channels $L^+ L^- \ra \gamma \gamma, \gamma Z, ZZ, Z H$, and $WW$, where $H$ is the Higgs boson (here we assume $M_{H} = 120$~GeV).

\section{FCHAMP Relic Abundance}
\label{relic}

The relic abundance of FCHAMPs may be expressed as a ratio of their mass density to the critical mass density~\cite{steigman1979} through the dimensionless mass density parameter $\Omega_{L} \equiv (\rho_{L}/\rho_{c})_{0}$, where $\rho_{L 0} = (\rho_{L^{+}} + \rho_{L^{-}})_{0} = 2m_{L}n_{L 0}$, and $\rho_{c 0}$, the present critical mass (energy) density, depends on the present value of the Hubble parameter, $H_{0} \equiv 100h$~km~Mpc$^{-1}$s$^{-1}$ ($h \approx 0.70$~\cite{freedman,komatsu}), $\rho_{c 0} = 1.05\times 10^{-5}h^{2}$~GeV/cm$^{3}$.  For the range of \Lpm~annihilation cross sections of interest here and, for $n_{\gamma 0} = 410.5$~cm$^{-3}$, corresponding to $T_{0} = 2.725$~K~\cite{mather},
\beq
\Omega_{L}h^{2} \approx {5.0\times 10^{-27} \over \langle \sigma v\rangle_{ann*}}.
\eeq
\label{oml1}
The thermally averaged annihilation rate factor, $\langle \sigma v\rangle_{ann}$, in units of cm$^{3}$s$^{-1}$, is evaluated using the techniques in~\cite{Gondolo:1990dk} at $T = T_{*} \approx m_{L}/25$, the temperature at which the abundance of the \Lpm~pairs begins to depart significantly from their equilibrium abundance~\cite{steigman1979}.  Note that the FCHAMPs are {\bf not} ``frozen-out" at $T = T_{*}$.  For $T \la T_{*}$ annihilations continue to dominate the evolution of \Lpm~pairs, further reducing their abundance.  Only when $T \ll T_{*}$ is the FCHAMP relic abundance ``frozen-out", although even then annihilations continue, albeit at a rate which is much smaller than the universal expansion rate, leaving the relic abundance effectively unchanged. 

The relic ratio of the number of \Lpm~pairs to CMB photons is
\beq
\bigg({n_{L} \over n_{\gamma}}\bigg)_{0} \approx {6.2\times 10 ^{-35} 
\over m_{L}({\rm GeV})\langle \sigma v\rangle_{ann*}}.
\eeq
Observations of the CMB temperature anisotropy spectrum constrain the value of the baryon (dominated by protons and alphas) to photon ratio, $n_{\rm B}/n_{\gamma} \approx 6.2\times 10^{-10}$~\cite{komatsu}.  Barring any significant post freeze-out renewed annihilation (see \S\ref{earth}), the present ratio of \Lpm~pairs to baryons is
\beq
\bigg({n_{L} \over n_{\rm B}}\bigg)_{0} \approx {1.0\times 10 ^{-25} \over m_{L}({\rm GeV})\langle 
\sigma v\rangle_{ann*}} = {1.0\times 10 ^{-27} \over m_{100}\langle \sigma v\rangle_{ann*}}.
\eeq
\label{nlnb}

In practice, we will evaluate  $\langle\sigma v\rangle_{ann}$ at threshold, i.e., $T \ll T_{*}$, in part because that is the relevant quantity for the subsequent annihilation in the Galaxy and on Earth. We have checked that there is little difference ($< 13$\%) in the rate factor between  $T = T_{*}$ and  $T \ll T_{*}$ (see Figure \ref{omhqedsm}).

\subsection{The Annihilation Rate Factor}

The total annihilation rate factor, $\langle\sigma v\rangle_{ann}$, the thermally averaged product of the annihilation cross section and the relative velocity, may be written in terms of the two-photon annihilation rate factor, $\langle\sigma v\rangle_{ \gamma \gamma}$, multiplied by $N$, a function of $Q_{L}$ and $m_{L}$ ($m_{L} \equiv 100\,m_{100}$ GeV), which accounts for all the open annihilation channels.  $N \equiv \langle\sigma v)_{ann}/(\sigma v\rangle_{ \gamma \gamma}$, where $\langle\sigma v\rangle_{ \gamma \gamma} = \pi  \alpha ^2 Q_L^4/m_L^2 = 2.2\times 10^{-25}Q_{L}^{4}/m_{100}^{2}$~cm$^{3}$s$^{-1}$ (we have adopted $\alpha = 1/128$).  In computing $N$ we have included, in addition to annihilation to two photons ($N_{\gamma\gamma} = 1$), the contributions from annihilations to all lepton and quark pairs (with $m < m_{L}$) along with contributions from annihilations to boson pairs $WW$, $WZ$, $ZZ$, $ZH$, and $Z\gamma$, where $H$ is the Higgs boson, for $m_{L} > (M_{B_1}+M_{B_2})/2$.  (We take $M_H=120$ GeV). $N$ may be written as $N=a+b/Q_L^2$, where $a$ results from annihilation into $\gamma\gamma$, $\gamma Z$, and $ZZ$, while $b$ represents annihilations  to fermion pairs (three generations of quarks and leptons) as well as into $WW$ and $ZH$. With these definitions,
\beq
\Omega_{L}h^{2} \approx {0.022m_{100}^{2} \over Q_{L}^{2}(aQ_{L}^{2} + b)}.
\eeq
\label{oml2}

\subsubsection{A Toy Model: Pure QED}

To understand (and anticipate) our results for $a = a(m_{L})$, $b = b(m_{L})$ and $\Omega_{L}h^{2}$ in the standard model (SM), it is useful to consider \Lpm~annihilation in a ``toy model", ``pure QED".  This model assumes that the FCHAMPs interact purely electromagnetically and that there are no $W$ or $Z$ bosons (or Higgs).  In this model the only annihilation channels are \Lpm~$\rightarrow \gamma\gamma$ and \Lpm~$\rightarrow f^{\pm}$.  In this case, $a^{QED} = 1$ and each fermion pair with $m_{f} > m_{L}$ contributes to $b^{QED}$ an amount $\propto q_{f}^{2}$.  For three generations of quarks and leptons, assuming that $m_{L} > m_{b}$ (or, $m_{L} \gg m_{b}$),  $b^{QED} = (20 + 4F(m_t/m_L))/3$ where, to account for the top quark threshhold at $m_{L} = m_{t}$,
\beq
F(x)\equiv \frac{1}{2} \sqrt{1-x^2} \left(x^2+2\right) \theta (1-x)\xrightarrow[x\ra 0]{} 1.
\label{mtqed}
\eeq

\begin{figure}[htbp]
\begin{center}
\includegraphics*[scale=0.45]{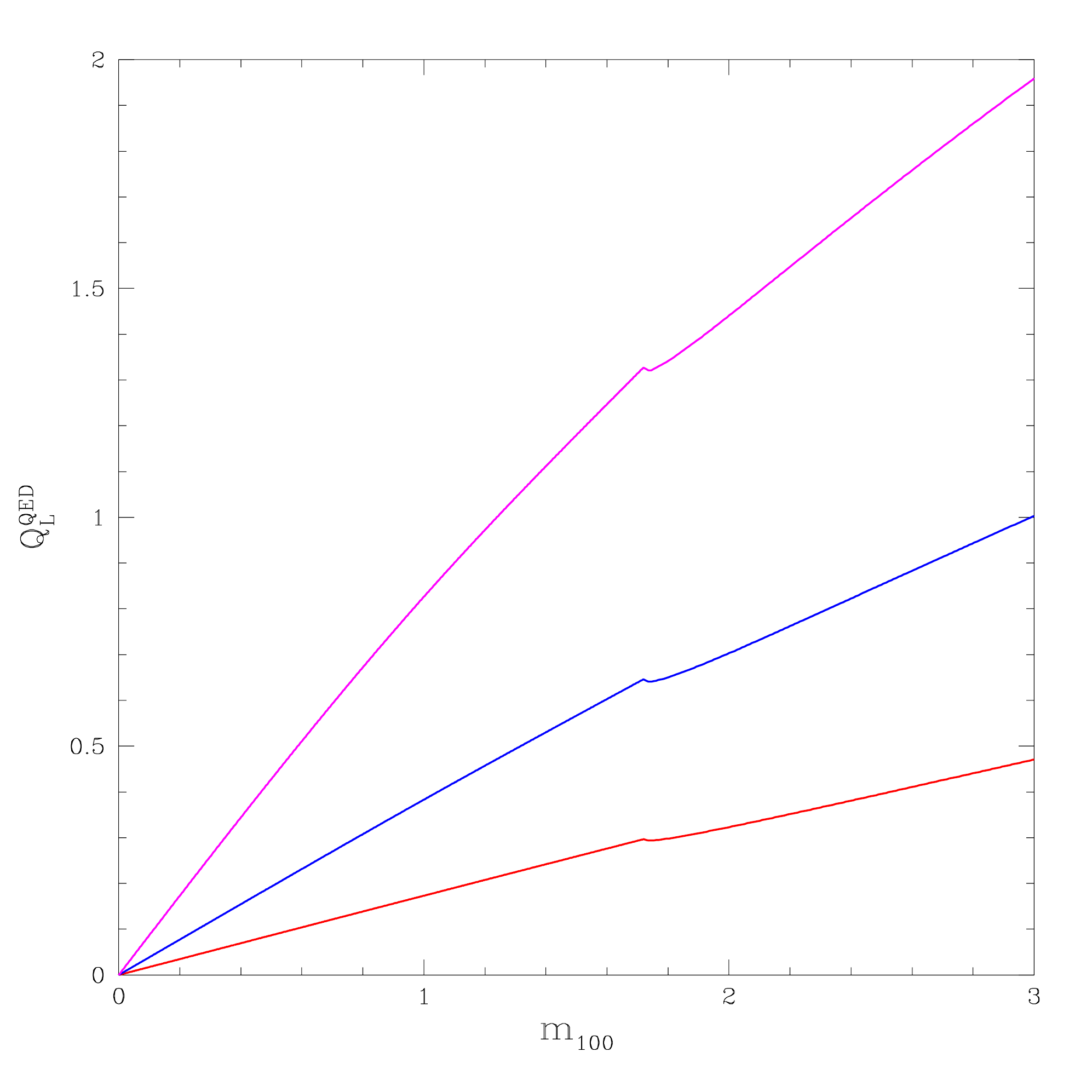}
\caption{Contours of fixed values of $\Omega_{L}$ in the $Q_{L} - m_{L}$ plane for the QED toy model.  The red curve corresponds to $\Omega_{L} = \Omega_{\rm C}$ (cold dark matter), the blue curve to $\Omega_{L} = \Omega_{\rm B}$ (baryons), and the magenta curve to $\Omega_{L} = \Omega_{\rm B}/5$; see the text for details.  The regions {\bf above} the curves are allowed.}
\label{qvsmqed}
\end{center}
\end{figure}

If \omegalh~is fixed, Eq.~4 leads to a quadratic equation for $Q_{L}^{2}$ as a function of $m_{L}^{2}$ which depends on $a = a(m_{L})$ and $b = b(m_{L})$.  The resulting \ql~-- \ml~relation for the pure QED toy model is shown for three choices of \omegal~in Figure~\ref{qvsmqed}.  The small ``glitches" at $m_{100} \approx 1.7$ reflect the enhanced annihilation cross section from the opening of the top quark channel\footnote{In principle there are also threshold effects for $m_L=m_b, m_c, m_\tau$, etc. These would be almost invisible on the scales of the figures and are not explicitly included. Our treatment of final state hadrons as free quarks should be a reasonable approximation for the ranges ($Q_{L} \ga 0.01$, $m_{L} \ga 1$~GeV) considered here.}.  A first, naive choice for a lower bound to \omegalh~is from the requirement that the FCHAMP relic density not exceed the density in cold dark matter, $\Omega_{L}h^{2} \la \Omega_{\rm C}h^{2} \approx 0.11$~\cite{komatsu}.  The region above the red curve in Figure~\ref{qvsmqed} satisfies this constraint.  However, charged, massive FCHAMPs do not behave like cold dark matter, they behave more like heavy ions or baryons~\cite{haibo}.  The region above the blue curve in Figure~\ref{qvsmqed} is consistent with $\Omega_{L}h^{2} \la (\Omega_{\rm B}h^{2})_{\rm BBN} \approx 0.022$~\cite{steigman2007}.  However, the relic abundances of the light elements, especially deuterium and helium-4, require this value of the relic density be provided by {\bf baryons}, so the limit on the FCHAMP density must be even smaller~\cite{rubtsov}.  Following the analysis of Dubovsky, Gorbunov and Rubtsov~\cite{rubtsov}, the magenta curve in Figure~\ref{qvsmqed} represents $\Omega_{L}h^{2} = \Omega_{\rm B}h^{2}/5 \approx 0.0044$.  It is the region above the magenta curve in Figure~\ref{qvsmqed} that is consistent with the density constraint; we use this constraint ($\Omega_{L}h^{2} \la 0.0044$, $\Omega_{L} \la \Omega_{\rm B}/5$) in our subsequent discussion.

\begin{figure}[htbp]
\begin{center}
\includegraphics*[scale=0.6]{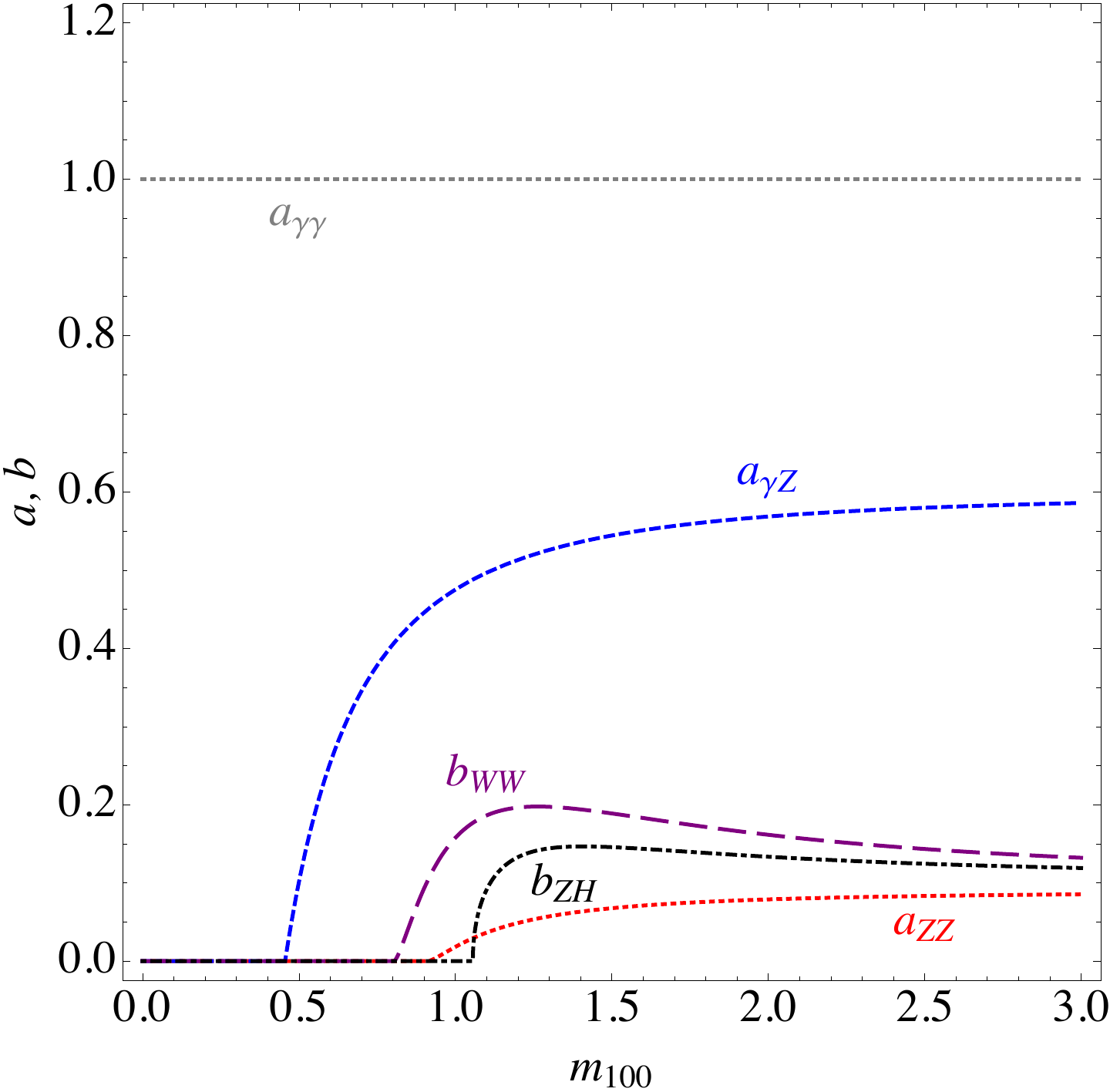}
\caption{The contributions to $a^{\rm SM}$ from annihilation to $\gamma\gamma$ (gray), $\gamma Z$ (blue),  and $ZZ$ (red), as well as the bosonic contributions to  $b^{\rm SM}$ from $WW$ (purple) and $ZH$ (black),  as a function of $m_{L} = 100\,m_{100}$~GeV.  Note that $a^{\rm QED} = a_{\gamma\gamma} = 1$.}
\label{bosonssm}
\end{center}
\end{figure}

\subsubsection{The Standard Model}

\begin{figure}[htbp]
\begin{center}
\includegraphics*[scale=0.6]{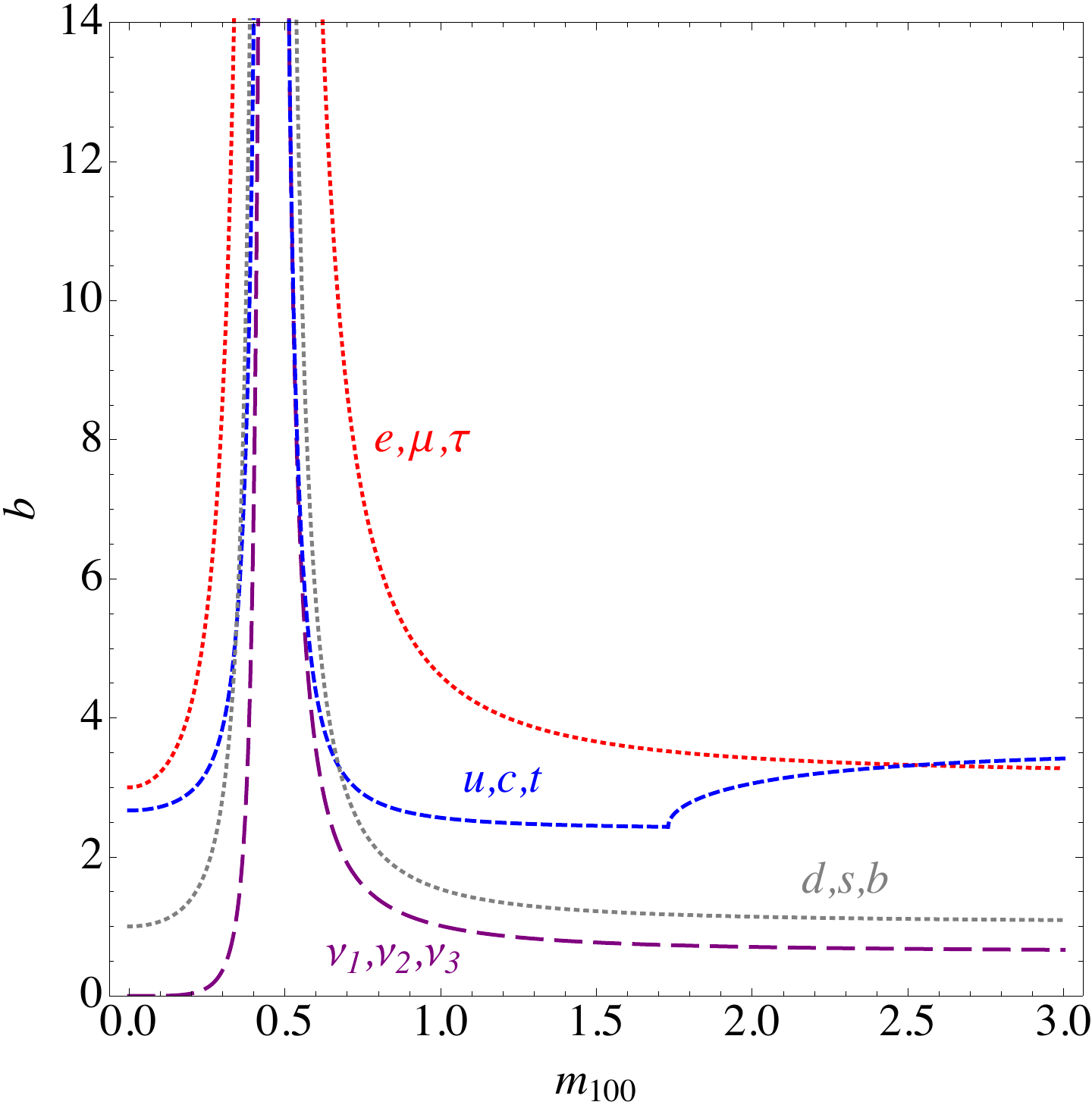}
\caption{The contributions to $b^{\rm SM}$ from annihilation to $e, \mu, \tau$ (red), $u, c, t$ (blue), $d, s, b$ (gray), and $\nu_i, i=1,2,3$ (purple) are shown as a function of $m_{L} = 100\,m_{100}$~GeV.}
\label{fermions}
\end{center}
\end{figure}

\begin{figure}[htbp]
\begin{center}
\includegraphics*[scale=0.45]{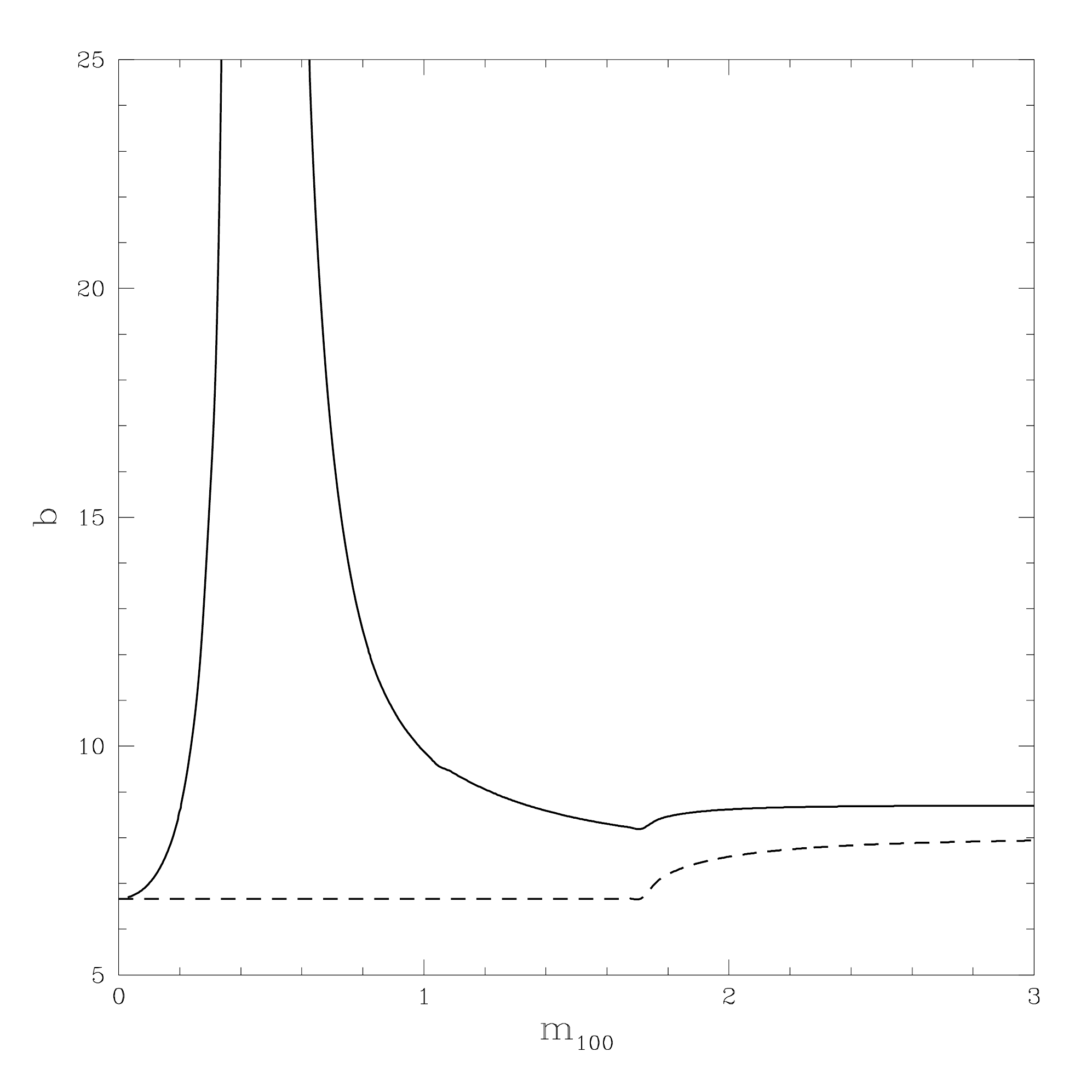}
\caption{The standard model (solid curve) and QED (dashed curve) values for $b$ from $L^{+}L^{-} \rightarrow f \bar f, W^{+}W^{-}, ZH$ as a function of $m_{L} = 100\,m_{100}$~GeV.}
\label{bvsm}
\end{center}
\end{figure}

With the pure QED toy model as background, we now return to the calculations of $a$, $b$, and the \ql~-- \ml~relation for the standard model (SM).  In addition to direct two photon annihilation, $a^{\rm SM}$ receives contributions from annihilation to $\gamma Z$ and to $Z$ pairs, as shown in Figure~\ref{bosonssm}.   The individual contributions to $b^{\rm SM}$ from annihilations into $WW$, $ZH$, and  fermion pairs  are shown in Figures~\ref{bosonssm} and \ref{fermions}. The total values for $b^{\rm SM}$ (solid curve) and $b^{\rm QED}$ (dashed curve) are shown in Figure~\ref{bvsm}, and the detailed formulae for the cross sections are given in the Appendix.  Notice the enormous enhancement due to the $Z$ resonance in the annihilation cross section for $m_{L} \sim M_{Z}/2$.

\begin{figure}[htbp]
\begin{center}
\includegraphics*[scale=0.6]{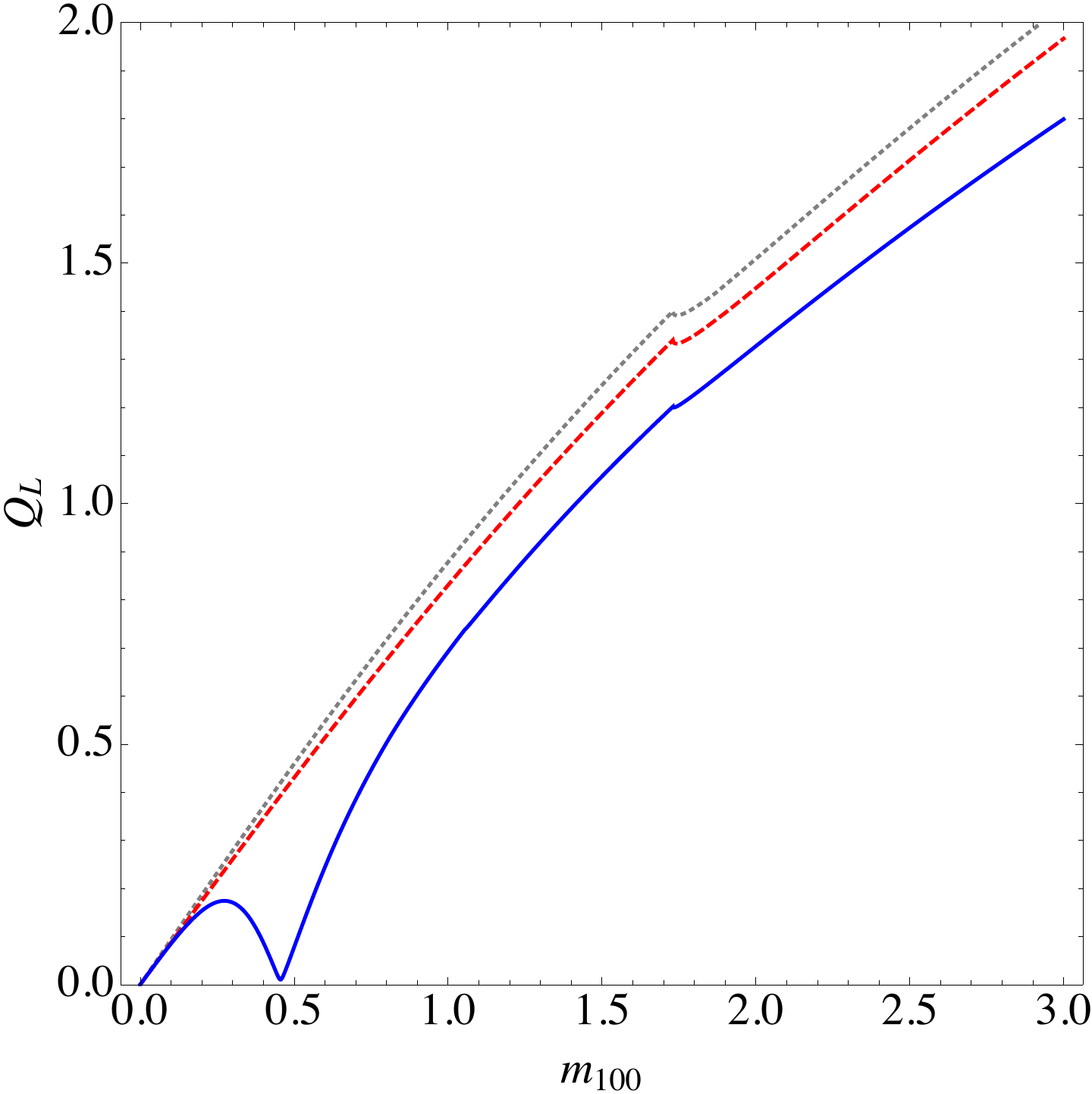}
\caption{Contours of $\Omega_{L} = \Omega_{\rm B}/5$ in the $m_{L} - Q_{L}$ plane for the QED toy model (red dashed curve) and the standard model (blue solid curve) using the annihilation rate factors calculated at threshold; see the text for details.  The regions {\bf above} the curves are allowed. For comparison, the gray dotted curve shows the constraint for the QED toy model using the annihilation rate factor evaluated at $T = T_{*} = m_{L}/25$.}
\label{omhqedsm}
\end{center}
\end{figure}

A comparison between the relic density constraints for the toy model (dashed curve) and the standard model (solid curve) which follow from the requirement that $\Omega_{L}h^{2} \la 0.0044$ are shown in Figure~\ref{omhqedsm}.  The enhanced SM annihilation cross section for $m_{L} \sim M_{Z}/2$ is responsible for the dip in the \ql~-- \ml~relation at low \ml.

\section{Accelerator and $Z$-Width Constraints on \ql~and \ml}
\label{acc}

\begin{figure}[htbp]
\begin{center}
\includegraphics*[scale=0.45]{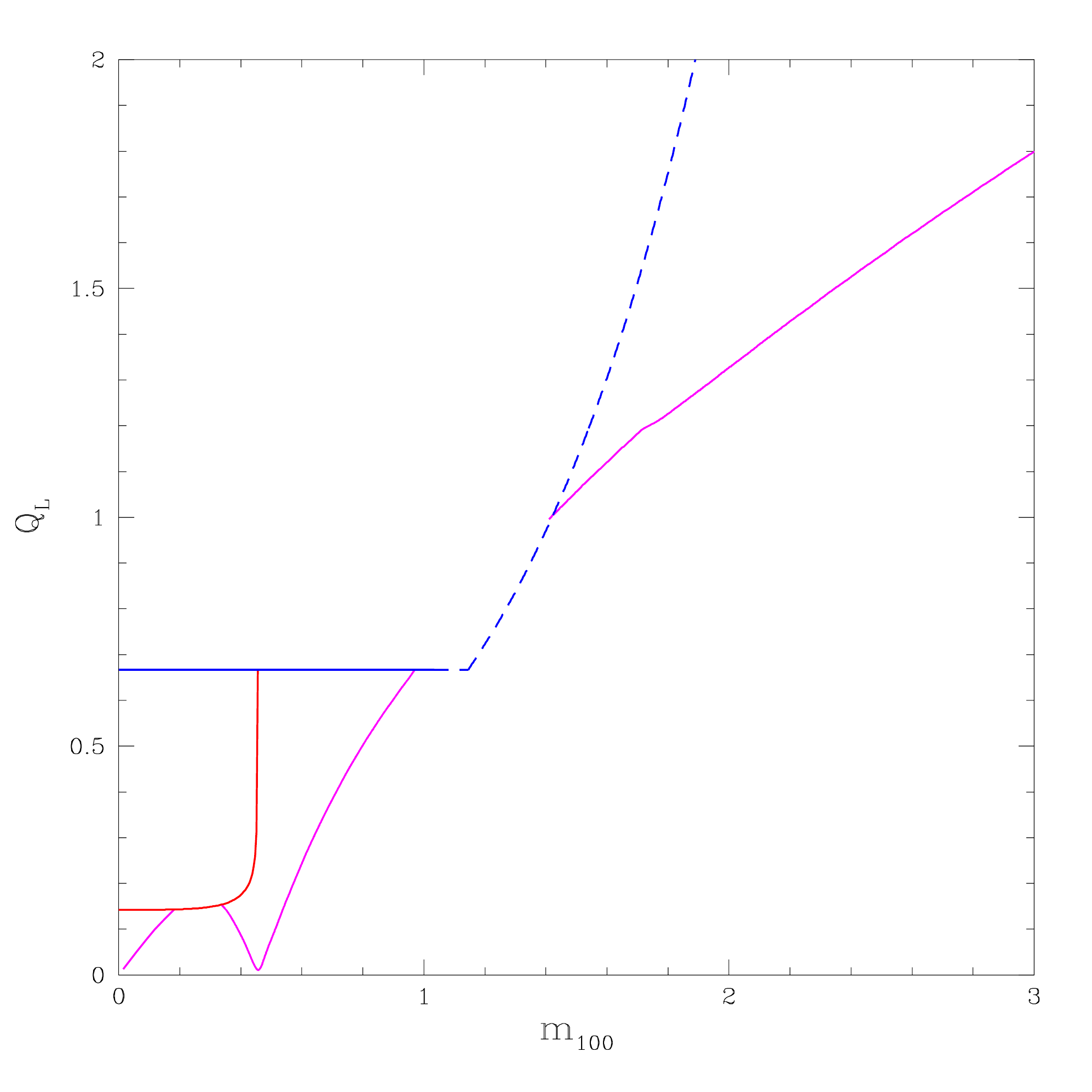}
\caption{Constraints in the $Q_{L} - m_{L}$ plane.  The magenta curve corresponds to $\Omega_{L} = \Omega_{\rm B}/5$ for the standard model.  The regions {\bf above} the magenta curve ($\Omega_{L} < \Omega_{\rm B}/5$) are allowed.  The region above the red curve is excluded by the constraint from the invisible $Z$-width.  The region above  the blue line at  \ql~= 2/3 and to the left of the dashed blue line is excluded by collider searches.  
See the text for details.}
\label{qvsmall1}
\end{center}
\end{figure}

The review article by Perl, Lee and Loomba~\cite{Perl:2009zz} provides a summary of the searches for fractionally charged elementary particles.  The best accelerator/collider constraint for a generic color-singlet fermion comes from the OPAL collaboration~\cite{abbiendi} at LEP 2, excluding $Q_{L} \geq 2/3$ for $m_{L} \leq 102$~GeV at 95\% c.l.  

The D0 collaboration at the Tevatron has set  95\% c.l. lower limits of 206 GeV and 171 GeV, respectively, on the
masses of long-lived charged gauginos and Higgsinos~\cite{Abazov:2008qu}.  For $m_{L} \la 300~{\rm GeV}$ these can be approximately translated into limits on unit-charged FCHAMPs by combining the theoretical NLO cross sections and experimental limits (as a function of mass) presented by D0 with the parton-level FCHAMP and gaugino Drell-Yan cross sections given in the Appendix. We find that $m_L \lesssim 142$ GeV for $Q_L=1$.  The result can be extended to other values of $Q_L$ if one makes the additional assumption that the acceptance depends only weakly on $Q_L$. We assume that this is valid for $Q_L \ge 2/3$, corresponding to $m_{L} \leq 115~{\rm GeV}$, in which case the combination of the OPAL and D0 collider limits excludes the region in the upper left corner of Figure~\ref{qvsmall1} bounded by the solid and dashed blue lines.

The $Z$ may decay to an FCHAMP pair provided that $m_{L} < M_{Z}/2$.  The limit to the invisible width of the $Z$, often expressed as a limit to the number of equivalent neutrinos, $N_{\nu} = 2.984 \pm 0.009$~\cite{nakamura}, provides a lower limit to $Q_{L} (\ga 0.16)$ as a function of \ml\footnote{This is the 95\% c.l. limit, taking into account that the physical value of $N_\nu-3 $ is positive semi-definite. The total $Z$-width obtained directly from the lineshape gives a  somewhat weaker lower limit, $Q_L\ga 0.24$ for $m_L < M_Z/2$.}.  This bound is shown as the red curve in Figure~\ref{qvsmall1}.

The regions above the magenta curves, below and to the right of the red curve, and below and to the right of the blue line and dashed curve in Figure~\ref{qvsmall1} identify the regions in the \ql~-- \ml~plane that are consistent with the relic density, collider, and invisible $Z$-width constraints.  We next consider the constraints imposed by terrestrial searches for fractionally charged, massive particles.

\section{Constraints from Terrestrial Searches for FCHAMPs}
\label{earth}

\begin{figure}[htbp]
\begin{center}
\includegraphics*[scale=0.45]{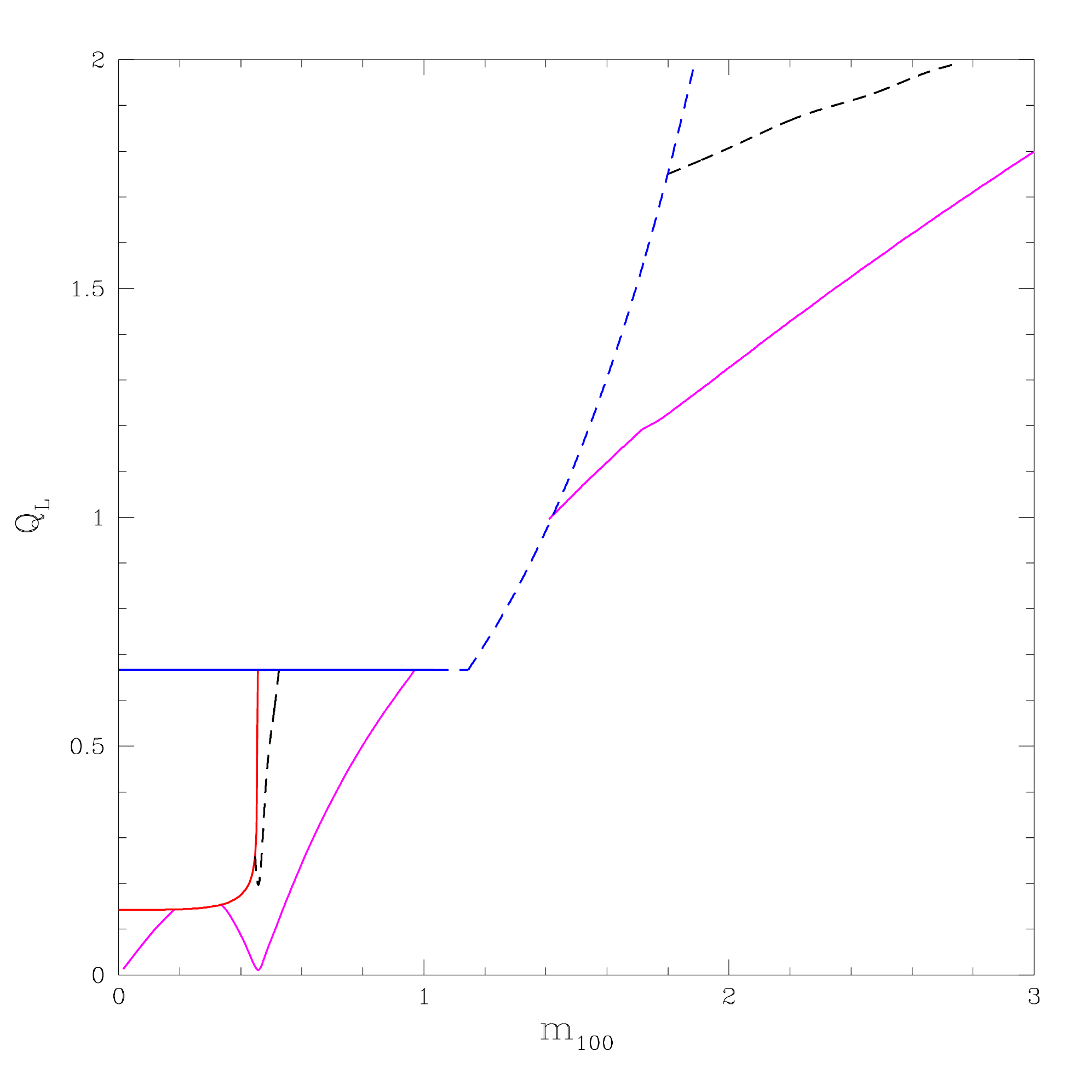}
\caption{As for Figure \ref{qvsmall1}, constraints in the $Q_{L} - m_{L}$ plane. The black, dashed curve corresponds to an allowed terrestrial FCHAMP abundance resulting from \Lpm~ annihilation over the Earth's history, {\bf provided} that all relic ${L^{-}}$ were {\bf free} (\ie not bound to alphas or protons).  The allowed region is {\bf above} the black, dashed curve (subject to the collider (blue) and invisible $Z$-width (red) constraints).  See the text for details.}
\label{qvsmall2}
\end{center}
\end{figure}

The null results of searches for fractionally charged particles in bulk matter on Earth (E)~\cite{kim2007,lee} or meteoritic material (\eg\cite{kim2007,jones}) are summarized in the review by Perl {\it et al.}~\cite{Perl:2009zz}.  Their 95\% confidence upper limits for terrestrial material from the oil drop or magnetic levitometer techniques, $(n_{L}/n_{\rm B})_{\rm E} < 1.3\times 10^{-23}$, and for meteoritic material (M), $(n_{L}/n_{\rm B})_{\rm M} < 7.0\times 10^{-22}$, provide enormously stringent limits to free terrestrial FCHAMPs\footnote{On average, the net FCHAMP charge on a drop would be zero, but there are fluctuations in the numbers of $L^{+}$ and $L^{-}$ in any drop (including those bound to protons and  alphas) and many of the fluctuations should yield drops whose net charge differs from an integer.}.  Since the FCHAMP density is $\rho_{L} = 2m_{L}n_{L}$ and the baryon density is $\rho_{\rm B} = m_{\rm B}n_{\rm B}$, the ratio of relic FCHAMPs to baryons by number may be written as
\beq {n_{L} \over n_{\rm B}} = {9.4\times 10^{-4} \over m_{100}}\bigg({\Omega_{L} \over \Omega_{\rm B}/5}\bigg) = {4.7\times 10^{-3}m_{100} \over Q_{L}^{2}(aQ_{L}^{2} + b)}.
\label{abundance}
\eeq
The relic abundance of FCHAMPs is many orders of magnitude larger than the upper bound to their density set by the terrestrial constraints.  However, long after annihilation freeze out in the early Universe, when the solar system and Earth form, annihilation resumes in the resulting high density, low temperature environment.  Indeed, at the low terrestrial temperature ($T \sim 300$~K), the annihilation cross section is ``Sommerfeld enhanced"~\cite{sommerfeld} from its early Universe value,
\beq
\langle\sigma v\rangle_{\rm E} = S(\eta_{\rm E})\langle\sigma v\rangle_{\rm E},
\eeq
where, for free \Lpm~pairs, $\eta_{\rm E} \equiv 2\pi Q_{L}^{2}(\alpha c/v)_{\rm E} \approx 7.2\times 10^{4}Q_{L}^{2}m_{100}^{1/2}$, $\langle\sigma v\rangle_{\rm E}$ is evaluated at threshhold ($T_{\rm E} \ll T_{*}$) and,
\beq
S(\eta) \equiv {\eta \over 1 - e^{-\eta}}\xrightarrow[\eta\gg 1]{} \eta.
\eeq
For equal numbers of positively and negatively charged FCHAMPs, renewed annihilation on Earth is rapid provided that $\Gamma_{ann}^{\rm E}\Delta t = n_{L}\langle\sigma v\rangle_{\rm E}\Delta t \gg 1$, where $\Delta t \approx 4.5$~Gyr and $n_{L} = n_{\rm B}(n_{L}/n_{\rm B}) \approx 6.4\times 10^{23}(n_{L}/n_{\rm B})$~\cite{Perl:2009zz}.  In this case, the annihilation on Earth efficiently reduces the relic $L$/B ratio to a present value of
\beq
\bigg({n_{L} \over n_{\rm B}}\bigg)_{\rm E} = {1 \over n_{\rm B}\langle\sigma v\rangle_{\rm E}\Delta t} = {6.9\times 10^{-22}m_{100}^{3/2} \over Q_{L}^{4}(aQ_{L}^{2} + b)}.
\eeq
The requirement that $(n_{L}/n_{\rm B})_{\rm E} < 1.3\times 10^{-23}$ leads to the constraint on \ql~as a function of \ml,
\beq
Q_{L}^{4}(aQ_{L}^{2} + b) > 53m_{100}^{3/2},
\eeq
which is shown by the dashed, black curves in Figure~\ref{qvsmall2}. 

The situation is quite different for free FCHAMPs in the ISM where $n_{\rm B} \approx 1$~cm$^{-3}$ and $T \approx 10^{4}$~K.  In this case, $\eta_{\rm ISM} \approx 1.25\times 10^{4}Q_{L}^{2}(m_{100}/T_{4})^{1/2}$ ($T \equiv 10^{4}T_{4}$~K) and the ISM annihilation rate for free FCHAMPs,
\beq
\Gamma_{ann}^{\rm ISM} = n_{L}\langle\sigma v\rangle_{\rm ISM} \approx 1.3\times 10^{-23}\bigg({Q_{L}^{2} \over m_{100}^{1/2}}\bigg)\bigg({n_{\rm B} \over T_{4}^{1/2}}\bigg),
\eeq
is far too small to deplete any free \Lpm~pairs in the Galaxy over the age of the Universe ($t \sim 14$~Gyr, $\Gamma_{ann}^{\rm ISM}t \ll1$).  In contrast to  free FCHAMPs on Earth, those in the ISM are effectively immortal.

\subsection{Constraints on \ql~And \ml}

As may be seen from Figure~\ref{qvsmall2}, for $Q_{L} < 2$ and $m_{L} < 300$~GeV, the only potentially allowed values of \ql~and \ml~are found above the dashed black curves, subject to the collider (blue) and invisible $Z$-width (red) constraints.  Although the regions above the relic abundance curve (magenta) are consistent with $\Omega_{L} < \Omega_{\rm B}/5$, in the regions below the dashed black curves the terrestrial abundance of FCHAMPs, ($n_{L}/n_{\rm B})_{\rm E}$, is too large.  For example, over the entire range of \ql~and \ml~shown in Figure~\ref{qvsmall2}, in the absence of further annihilation, the relic FCHAMP to baryon ratio exceeds $\sim 5\times 10^{-7}$.  In the regions above the dashed black curves annihilation on Earth can reduce this ratio to $< 1.3\times 10^{-23}$, consistent with the terrestrial constraints.  However, if some FCHAMPs, even a tiny fraction of them, are insulated from annihilation, even these allowed values of \ql~and \ml~values may be excluded by the terrestrial constraints.  We address this issue next.

\section{Bound States of $L^{-}$ with Alpha Particles and Protons}
\label{capture}

As the early Universe expands and cools, annihilation of free \Lpm~pairs effectively ceases but, at sufficiently low temperatures in the early Universe, after primordial nucleosynthesis (big bang nucleosynthesis: BBN) has converted some nucleons into alpha particles, collisions between alphas and protons and the negatively charged FCHAMPs ($L^{-}$) produce bound $L^{-}\alpha$, and $L^{-}p$ systems ($\equiv Lp$, $L\alpha$)~\cite{derujula,dimopoulos,gould,davidson} with charges $2 - Q_{L}$ and $1 - Q_{L}$ respectively.  For $Q_{L} > 1$ (2), $Lpp$ ($L\alpha\alpha$) bound states with electric charges $2 - Q_{L}$ ($4 - Q_{L}$) will form, replacing the $Lp$ ($L\alpha$) bound states.  Because of the exceedingly large ratio of cosmic background photons to FCHAMPs, these bound states can't form until $T \ll E_{b}$, typically not until $T \la E_{b}/40$, where $E_{b}$ is the binding energy of the system (or, of the system in its first excited state)~\cite{peebles}.  Since for the $L\alpha$ and $Lp$ bound states this occurs after BBN has ended ($T_{\rm BBN} \ga 30$~keV), these bound states will not affect the BBN-predicted relic abundances (unless $Q_{L} \ga 2$).  However, since there are considerably more baryons (nucleons) than FCHAMPs, a significant fraction of all $L^{-}$ will emerge from the early Universe ``dressed" by alphas or protons in tightly bound, positively charged, heavy ions.  Collisions between the free $L^{+}$ and these positively charged heavy ions are ``Gamow suppressed"~\cite{gamow} by their mutual Coulomb repulsion (the flip side of Sommerfeld enhancement between oppositely charged particles).

\begin{figure}[htbp]
\begin{center}
\includegraphics*[scale=0.45]{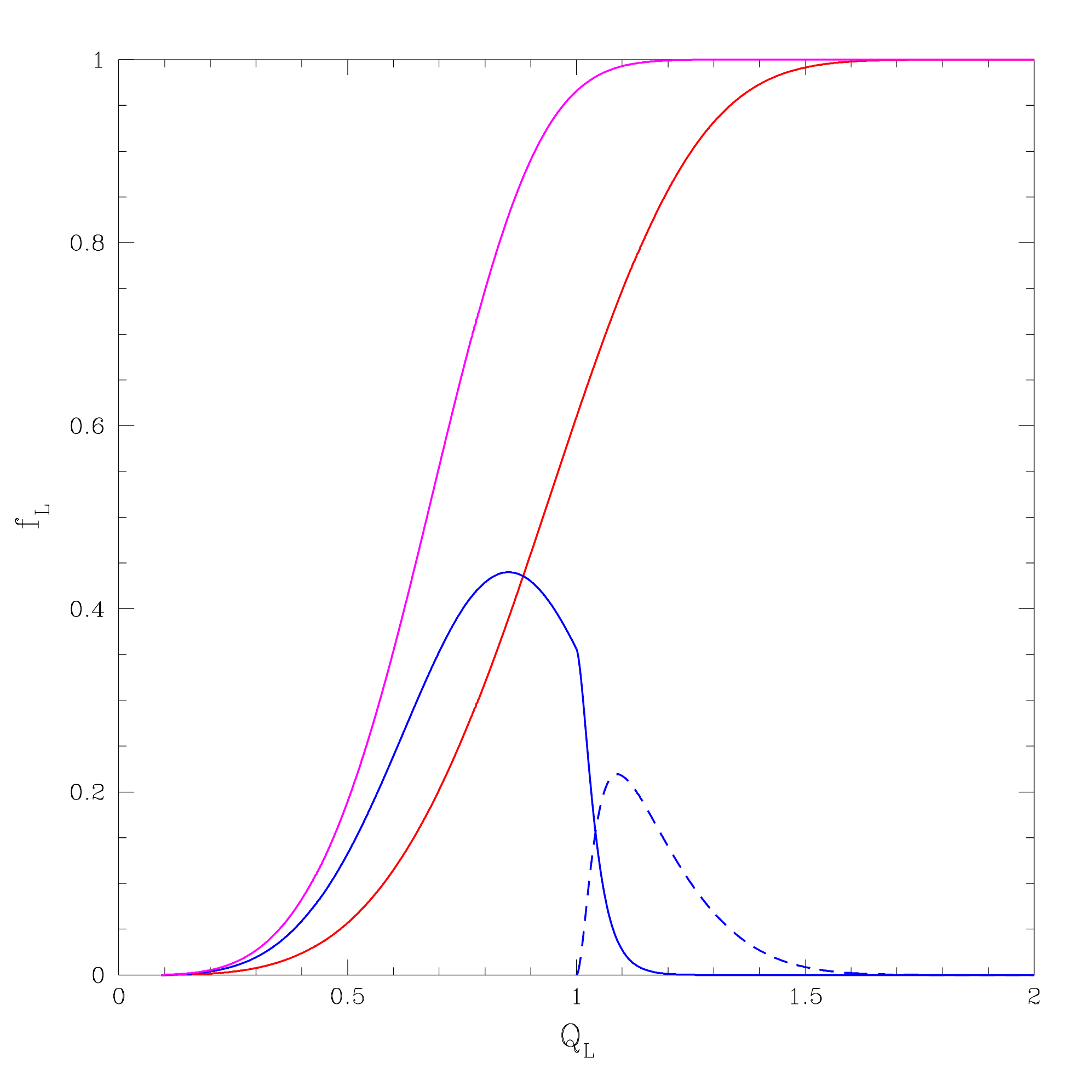}
\caption{The fraction of negatively charged FCHAMPs, $f_{L}$, captured into bound states with alphas (red), with protons (solid blue) and with two protons (dashed blue) as a function of \ql.  The magenta curve shows the sum.}
\label{fcaptot}
\end{center}
\end{figure}

\subsection{Capture of $L^{-}$ by Alpha Particles}

After BBN has ended the fraction of all baryons in alpha particles is $n_{\alpha}/n_{\rm B} =$ Y$_{\rm P}/4 \approx 1/16$~\cite{steigman2007}, where Y$_{\rm P}$ is the primordial helium mass fraction.  Using the estimate in eq.~6 for the baryon to FCHAMP ratio,
\beq
{n_{L} \over n_{\alpha}} \approx {0.015 \over m_{100}}\bigg({\Omega_{L} \over \Omega_{\rm B}/5}\bigg) < {0.015 \over m_{100}}.
\eeq
In the early Universe alpha particles outnumber FCHAMPs by a large factor.  At sufficiently low temperature, $T \la E_{b}(L\alpha)/40$, free $L^{-}$ will combine with alpha particles to form hydrogen-like bound states.  For $Q_{L} < 2$, the $L\alpha$ bound state has a net positive electric charge $2 - Q_{L} > 0$, a binding energy $E_{b}(L\alpha) \approx 400Q_{L}^{2}$~keV, and a ground state radius $r(L\alpha) \approx 3.6/Q_{L}$~fm\footnote{Since the radius of the alpha particle is comparable to the ground state radius of the $L\alpha$ ion, the $L^{-}$ will not see the full charge of the alpha particle and there are small corrections to the binding energy and radius which are ignored here.  The difference between the alpha particle mass and the reduced mass of the $L\alpha$ system has also been ignored.}.  This bound state forms when $T \la 10Q_{L}^{2}~{\rm keV}$ ($\la 40$~keV for $Q_{L} < 2$), after the alpha particles have formed during BBN. 

However, just because there are more alpha particles than FCHAMPs and $L\alpha$ bound states can form, does not guarantee that they will form.  An estimate of the fraction of negatively charged FCHAMPs captured by alpha particles may be found by comparing $L\alpha$ formation to ordinary proton -- electron (re)combination, which occurs when the temperature drops below $\sim E_{b}(H)/40 \approx 0.3$~eV.  At this time the recombination rate, $\Gamma_{pe} = n_{e}\langle\sigma v \rangle_{pe}$, greatly exceeds the universal expansion rate, $H$,  $(\Gamma_{pe}/H)_{rec} \approx 2.6\times 10^{3}$~\cite{peebles}.  The $L\alpha$ (re)combination rate, $\Gamma_{L\alpha,rec} \equiv (n_{\alpha}\langle v\sigma \rangle_{L\alpha})_{rec}$ may be scaled to the $pe$ recombination rate, since
\beq
\bigg({n_{\alpha} \over n_{e}}\bigg)_{rec} \approx {1 \over 12}\bigg({T_{L\alpha} \over T_{pe}}\bigg)^{3}_{rec} \approx 2.1\times 10^{12}Q_{L}^{6},
\eeq
where $T_{L\alpha,rec}/T_{pe,rec} = E_{b}(L\alpha)/E_{b}(pe) = 4Q_{L}^{2}(m_{\alpha}/m_{e}) \approx 2.9\times 10^{4}Q_{L}^{2}$,
and
\begin{multline}
\bigg({\langle \sigma v \rangle_{L\alpha} \over \langle \sigma v \rangle_{pe}}\bigg)_{rec} = (2Q_{L})^{3}\bigg({m_{e} \over m_{\alpha}}\bigg)^{3/2}\bigg({T_{pe} \over T_{L\alpha}}\bigg)^{1/2}_{rec}\\ = (2Q_{L})^{6}\bigg({T_{pe} \over T_{L\alpha}}\bigg)^{2}_{rec} \approx 7.4\times 10^{-8}Q_{L}^{2},
\end{multline} 
so that
\beq
\bigg({\Gamma_{L\alpha} \over \Gamma_{pe}}\bigg)_{rec} = {(2Q_{L})^{6} \over 12}\bigg({T_{L\alpha} \over T_{pe}}\bigg)_{rec} \approx 1.6\times 10^{5}Q_{L}^{8}.
\eeq
Since $pe$ (re)combination occurs close to when the Universe is making the transition from radiation dominated to matter dominated,
\beq
(H_{pe}/H_{0})_{rec}^{2} = \Omega_{\rm R}(1 + z_{pe,rec})^{4} + \Omega_{\rm M}(1 + z_{pe,rec})^{3}.
\eeq
For $\Omega_{\rm M}/\Omega_{\rm R} \equiv 1 + z_{eq} \approx 3.2\times 10^{3}$~\cite{komatsu} and $1 + z_{pe,rec} \approx 1.1\times 10^{3}$, $H_{pe,rec} \approx 4.9\times 10^{-14}$~s$^{-1}$, so that $(\Gamma/H)_{pe,rec} \approx 2.6\times 10^{3}$.  In constrast, $L\alpha$ (re)combination occurs when the Universe is completely radiation dominated so that $H_{L\alpha,rec}/H_{0} = \Omega_{\rm R}^{1/2}(1 + z_{L\alpha,rec})^{2}$, where $(1 + z_{L\alpha,rec})/(1 + z_{pe,rec}) = T_{L\alpha,rec}/T_{pe,rec}$, so that
\beq
\bigg({(\Gamma_{L\alpha}/H) \over (\Gamma_{pe}/H)}\bigg)_{rec} \approx 10.6Q_{L}^{6}\bigg({T_{pe} \over T_{L\alpha}}\bigg)_{rec} \approx 3.6\times 10 ^{-4}Q_{L}^{4},
\eeq
and $(\Gamma_{L\alpha}/H)_{rec} \approx 0.94Q_{L}^{4}$.  Unlike $(\Gamma/H)_{pe,rec}$ which is $\gg 1$, $(\Gamma_{L\alpha}/H)_{rec}$ increases from $\ga 9.4\times 10^{-9}$ to $\la 15$ as $Q_{L}$ increases from $\ga 0.01$ to $\la 2$.  An estimate of the fraction of FCHAMPs captured by alpha particles is given by $f_{L\alpha} = [1 - {\rm exp}(-(\Gamma_{L\alpha}/H)_{rec})]$~\cite{dimopoulos} which, for $0.01 \la Q_{L} < 2$, increases by some eight orders of magnitude from $\sim 10^{-8}$ to 1.0; see Figure~\ref{fcaptot}.  Although for parts of this \ql~range most of the negatively charged FCHAMPs may escape capture by alpha particles leaving them vulnerable to annihilation, a non-negligible fraction of them will be sequestered in tightly bound, positively charged $L\alpha$ ions, insulating them from annihilation with their free, positively charged counterparts.  This leads to a ratio of the surviving FCHAMPs to baryons on Earth, $f_{L}(n_{L}/n_{\rm B})$, far in excess of the limits set by the terrestrial searches ($\la 1.3\times 10^{-23}$)~\cite{Perl:2009zz}. 

While the evolutionary history of meteoritic material is uncertain, such samples may represent a bridge from interstellar to terrestrial matter.  Since in the ISM $\Gamma_{ann}\Delta t \ll 1$, even for free FCHAMPs, the interstellar FCHAMP to baryon ratio is the relic value (see eq.~6), far exceeding the limits set by searches for fractionally charged particles terrestrially or in the meteorites ($\la 7.0\times 10^{-22}$)~\cite{Perl:2009zz}.

\subsection{Capture of $L^{-}$ By Protons}

Since the fraction of all baryons in the early, post-BBN Universe which are free protons is $n_{p}/n_{\rm B} = 1 - $Y$_{\rm P} \approx 0.75$~\cite{steigman2007}, the FCHAMP to proton ratio, by number, is
\beq
{n_{L} \over n_{p}} \approx {1.25\times 10^{-3} \over m_{100}}\bigg({\Omega_{L} \over \Omega_{\rm B}/5}\bigg) < {1.25\times 10^{-3} \over m_{100}}.
\eeq

The $Lp$ bound state forms later than the $L\alpha$ ion, when $T \la T_{Lp,rec} \approx 0.5Q_{L}^{2}$~keV.  Any $L^{-}$ which had failed to form $L\alpha$ ions may combine with protons to form $Lp$ bound states at this lower temperature.  For $Q_{L} < 1$, the $Lp$ ion has a net positive electric charge $1 - Q_{L} > 0$, a binding energy $E_{b}(Lp) \approx 25Q_{L}^{2}$~keV, and a ground state radius $r(Lp) \approx 29/Q_{L}$~fm.

As was the case for the capture of $L^{-}$ by alpha particles, it must be asked what fraction of any free $L^{-}$, those which escaped earlier capture by alpha particles, will be captured by protons when $T \la T_{Lp,rec}$.  Scaling the above analysis for $L\alpha$ to the $Lp$ system results in $(\Gamma_{Lp}/H)_{rec} \approx 2.43Q_{L}^{4}$.  The corresponding $Lp$ capture fraction, 
\beq
f_{Lp}(1 - f_{L\alpha}) = (1 - {\rm exp}(-(\Gamma/H)_{Lp,rec})){\rm exp}(-(\Gamma/H)_{L\alpha})
\eeq 
is shown in Figure~\ref{fcaptot} (solid blue curve).

The fraction of all negatively charged FCHAMPs which are either captured by protons or by alpha particles, 
\begin{multline}
f_{L tot} = 1 - {\rm exp}(-(\Gamma/H)_{Lp,rec} - (\Gamma/H)_{L\alpha,rec})\\ = 1 - {\rm exp}(-3.37Q_{L}^{4}), 
\end{multline}
ranges from $\ga 5.4\times 10^{-3}$ to $\la 0.97$ as $Q_{L}$ increases from 0.2 to 1 (see Fig.~\ref{fcaptot}).  

It is likely that later in the evolution of the Universe, \eg in the galactic interstellar medium or, perhaps in stars, strongly exothermic charge transfer reactions $Lp + \alpha \rightarrow L\alpha + p$ will rearrange all the bound, negatively charged FCHAMPs into $L\alpha$ heavy ions with charge $2 - Q_{L} > 1$.  

\subsubsection{$Lpp$ Bound States}

For $Q_{L} > 1$, the $Lp$ bound state is negatively charged, which would result in an {\bf enhancement} of the annihilation with free $L^{+}$.  However, when $Q_{L} > 1$, two-proton bound states with $L^{-}$ become possible.  Scaling to the binding energy of the $H^{-}$ system (replacing the proton with an $L^{-}$ and the two electrons with two protons), for $Q_{L} \ga 1$, $E_{b}(Lpp) \ga 1.4$~keV.  Repeating the above analysis for the formation of the $Lpp$ bound state leads to the dashed blue curve in Figure~\ref{fcaptot}.  For $Q_{L} \ga 1$, the $Lp$ bound states are ``replaced" by $Lpp$ bound states with positive charge 2 -- \ql~(for $Q_{L} < 2$).

\subsubsection{$L\alpha\alpha$ Bound States}

The analysis here has been limited to $Q_{L} < 2$.  For $Q_{L} > 2$ the $L\alpha$ bound state has negative charge and this might lead to an enhancement in the late time annihilation of \Lpm~pairs.  However, in analogy with the formation of the $Lpp$ bound state when $Q_{L} > 1$, a two-alpha bound state can form with $L^{-}$ when $Q_{L} > 2$.  Scaling to the binding energy of the neutral helium atom (replacing the helium nucleus with an $L^{-}$ and the two electrons with two alpha particles), for $Q_{L} > 2$, $E_{b}(L\alpha\alpha) \ga 180$~keV.  For $Q_{L} \ga 2$, the $L\alpha$ bound states would be ``replaced" by $L\alpha\alpha$ bound states with positive charge 4 -- \ql~(for $Q_{L} < 4$).  We note that for $Q_{L} > 2$, $L\alpha$ (re)combination would occur for $T_{rec} \ga 40$~keV, during BBN (after the formation of alpha particles).  It is not unlikely that these negatively charged particles might alter the BBN abundances of nuclides heavier than helium by helping to bridge the gaps at mass-5 and mass-8, perhaps leading to the BBN formation of $L^{-}$ bound states with C, N, O nuclei which would have even larger positive charge.  Investigation of this possibility is beyond the scope of this paper.

\section{Discussion}
\label{discussion}

Consider the consequences of the capture of $L^{-}$ by alpha particles and protons in the early Universe.  Over the range $0.01 \la Q_{L} \leq 2/3$ ($1 \la m_{L} \leq 115$~GeV), the fraction of the relic, negatively charged FCHAMPs in positively charged bound states with alpha particles or protons increases from $\sim 3\times 10^{-8}$ to $\sim 0.49$.  Even though this leaves most of the relic $L^{-}$ free to annihilate on Earth with the relic $L^{+}$, the bound $L^{-}$ are insulated from annihilation.  For $m_{L} \leq 115$~GeV, the surviving ratio of FCHAMPs to baryons on Earth, $f_{L}(n_{L}/n_{\rm B})$, exceeds $\sim 9\times 10^{-11}$, more than 12 orders of magnitude higher than the terrestrial upper limit.  All combinations of \ql~and \ml~in this regime are excluded.  The conflict is even worse at higher masses.  For $142 \leq m_{L} \leq 300$~GeV, the D0 constraint~\cite{Acosta:2002ju} provides an {\bf upper} bound to the charge which increases with \ml, from $Q_{L,max} = 1.0$ at $m_{L} = 142$~GeV to $Q_{L,max} = 10.3$ at $m_{L} = 300$~GeV, leading to a {\bf lower} bound to the relic FCHAMP to baryon ratio which decreases from $n_{L}/n_{\rm B} \ga 6\times 10^{-4}$ at \ml~= 142 GeV to $n_{L}/n_{\rm B} \ga 7\times 10^{-7}$ at \ml~= 300 GeV.  For $m_{L} \ga 300~{\rm GeV}$ there is no upper bound to \ql~so that for fixed \ml~the lower bound to the relic FCHAMP to baryon ratio decreases as a power of $Q_{L}$, $(n_{L}/n_{\rm B})_{min} \propto Q_{L}^{-4}$.  However, to reduce the FCHAMP abundance on Earth to a level consistent with the terrestrial upper bound would require a value of \ql~far in excess of the range considered here.

\subsection{Prying Open The Window For FCHAMPs}

\begin{figure}[htbp]
\begin{center}
\includegraphics*[scale=0.45]{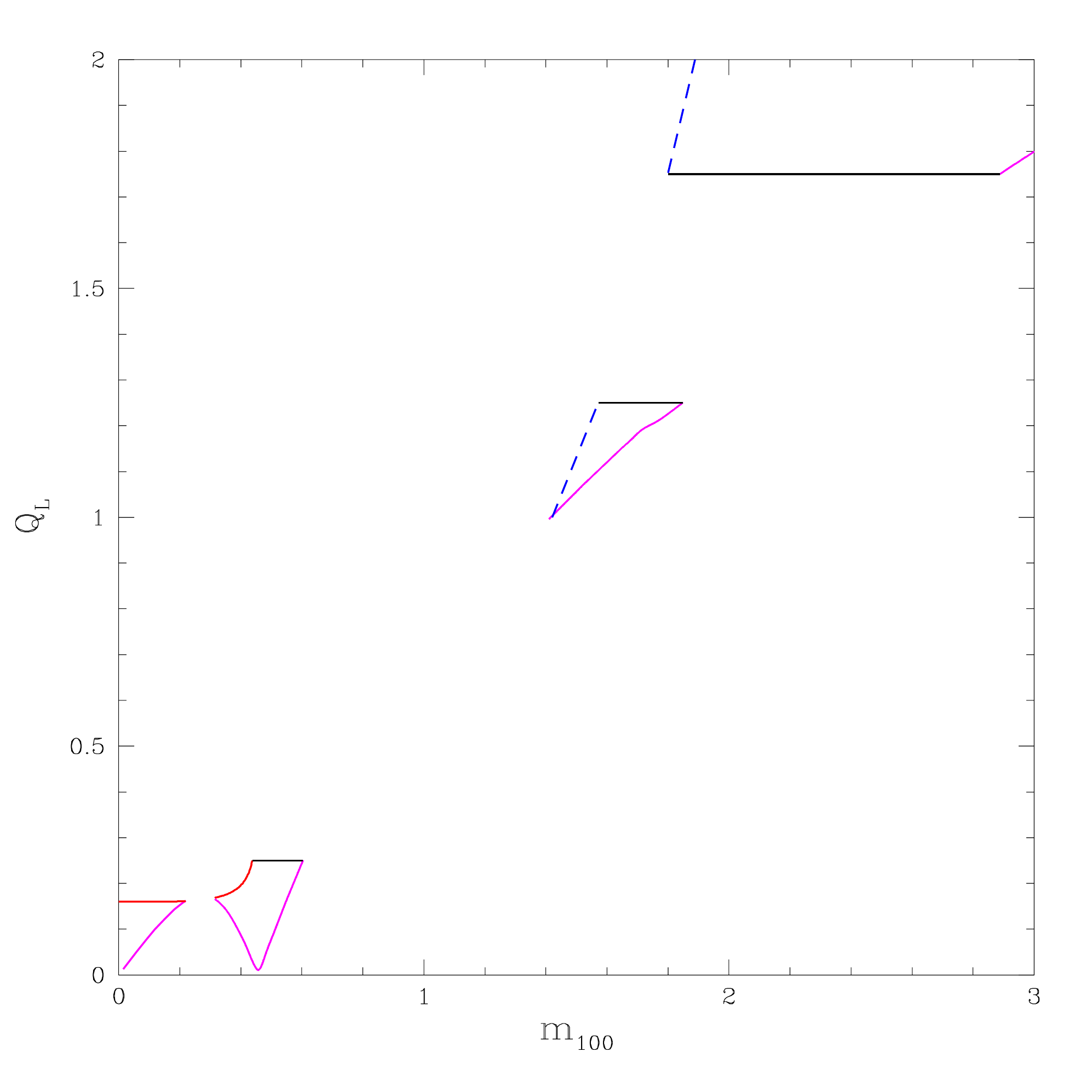}
\caption{Islands in the \ql~-- \ml~plane consistent with the relic density, accelerator, and $Z$ width constraints, for which $|Q_{L} - n| \leq 0.25$ ($n = 0, 1, 2$).}
\label{islands}
\end{center}
\end{figure}

The very strong conclusions reached here, precluding the existence of FCHAMPs with $Q_{L} \ga 0.01$ ($m_{L} \ga 1$~GeV), follow from the very low upper bounds to the FCHAMP to baryon ratio on Earth inferred from many oil drop and magnetic levitometer searches using terrestrial material (see~\cite{Perl:2009zz} and references therein).  However, these searches become insensitive to FCHAMPs as \ql~approaches an integer.  According to Perl \etal~\cite{Perl:2009zz}, FCHAMPs may not have been excluded if $|Q_{L} - n| \la 0.25$, where $n$ is an integer.  Islands in the \ql~-- \ml~plane satisfying $|Q_{L} - n| \leq 0.25$ and consistent with the relic density ($\Omega_{L} \leq \Omega_{\rm B}/5$), accelerator, and $Z$ width constraints are shown in Figure~\ref{islands}.  

If, indeed, the combinations of \ql~and \ml~shown in Figure~\ref{islands} have not been excluded by searches in terrestrial material, such FCHAMPs may be observable in the cosmic rays.  The cosmic rays result from the acceleration of material in the ISM of the Galaxy and if the interstellar FCHAMPs are efficiently accelerated\footnote{Note that FCHAMPs have a much smaller charge-to-mass ratio than do the ordinary nucleonic cosmic rays which may affect the acceleration efficiency.}, the resulting cosmic ray FCHAMP to nucleon ratio might be detectable.  Since free, as well as bound, ISM FCHAMPs avoid annihilation, the predicted distribution of fractionally charged particles in the cosmic rays provides a smoking gun for FCHAMPs.  For example, depending on the value of \ql, there will be free $L^{\pm}$ with $q = \pm Q_{L}$, along with $L\alpha$ with $q = +(2 - Q_{L})$ and $Lp$ with $q = +(1 - Q_{L})$ (and/or $Lpp$ with $q = +(2 - Q_{L})$ for $Q_{L} > 1$ and $L\alpha\alpha$ with $q = +(4 - Q_{L})$ for $Q_{L} > 2$).  The relative abundances of these charge states is determined by the relative capture fractions, $f_{L\alpha}$, $f_{Lp}$ ($f_{Lpp}$, $f_{L\alpha\alpha}$), and $f_{tot}$ (see Figure~\ref{fcaptot}).

To illustrate the potential of the cosmic rays to provide a smoking gun for FCHAMPs, consider the following example.  Suppose that \ql~= 5/6 ($|Q_{L} - 1| = 1/6 < 0.25$) and \ml~= 110 GeV.  For this choice of parameters, $\Omega_{L} = 0.17\Omega_{\rm B} < \Omega_{\rm B}/5$ and the relic FCHAMP to baryon ratio by number is $n_{L}/n_{\rm B} = 7\times 10^{-4}$.  For \ql~= 5/6,  the fraction of negatively charged FCHAMPs that are free ($q= -5/6$) is $1 - f_{tot} = 0.20$ so that $N(-5/6)/N(+5/6)=0.20$, $N(+7/6)/N(+5/6) = f_{L\alpha} = 0.36$ and, $N(+1/6)/N(+5/6) = f_{Lp} = 0.44$.

\section{Conclusions}
\label{conclusions}

FCHAMPs with $Q_{L} \ga 0.01$ and $m_{L} \ga$~few GeV behave like heavy baryons~\cite{haibo}, leading to an interesting constraint on their relic mass density~\cite{rubtsov}, $\Omega_{L} \la \Omega_{\rm B}/5$ ($\Omega_{L}h^{2} \la 0.0044$).  However, if this constraint were saturated the relic FCHAMP abundance on Earth would be orders of magnitude larger than the limits set by the negative results of searches for terrestrial, fractionally charged particles.  But, if the relic terrestrial FCHAMPs consisted of ``free" \Lpm~pairs, renewed, Sommerfeld enhanced, annihilation would have occurred over the past 4.5 Gyr in the high density, low temperature environment of the Earth, reducing the relic abundance and resulting in a range of \ql~-- \ml~values ($Q_{L} \ga 0.2$, $m_{L} \ga M_{Z}/2$; see Figure~\ref{qvsmall2}) consistent with the upper bounds inferred from the terrestrial searches.  However, as we have noted (see \S\ref{capture}), in the early Universe after \Lpm~freeze out, negatively charged FCHAMPs will be captured into positively charged bound states with alpha particles and protons, suppressing any further, late time annihilation.  This would appear to close the window on {\bf any} FCHAMPs with $0.01 \la Q_{L} <2~(< 4)$.  Although a ``natural" upper bound for the range of the electric charge of an FCHAMP might be $Q_{L} < 1$, we have extended the analysis here to $Q_{L} < 2$ (or, to $Q_{L} < 4$ by including $L\alpha\alpha$). We expect that larger values of $Q_L$ would also be excluded by the negative results of searches for fractionally-charged particles on Earth because of bound states with multiple $\alpha$'s and/or heavier nuclei.  In that case, Eq.~\ref{abundance} would suggest that only enormous values of $Q_L (> 10^4-10^5$) would be allowed.  We have not considered the very large $Q_L$ scenario here because it would likely require modifying BBN and, furthermore, perturbative calculations would break down for $\alpha Q_L^2/4 \pi =\mathcal{O}(1)$.  Note that for \ml~$\la 300$~GeV, the D0 constraint~\cite{Acosta:2002ju} suggests that $Q_{L,max} \la 10.3$, resulting in a {\bf lower} bound to the relic FCHAMP to baryon ratio, $n_{L}/n_{\rm B} \ga 7\times 10^{-7}$.

Having argued that the extremely low upper bound from terrestrial searches for FCHAMPs has closed the window on such particles, we then noted that such searches may have been insensitive to FCHAMPs whose charge is close to an integer ($|Q_{L} - n| \leq 0.25$) and we identified some islands in the \ql~-- \ml~plane (see Figure~\ref{islands}) where fractionally charged massive particles may not have been excluded.  FCHAMPs with these charges and masses would reside in the ISM of the Galaxy and may have been incorporated into the cosmic rays.  Searches for such particles and, if there is evidence for them, the ratios of different charge states ($\pm Q_{L}$, $+(2 - Q_{L})$, $+(1 - Q_{L})$, etc.) can provide a smoking gun for the existence of FCHAMPs.  It is also anticipated that the collider limits, especially for large \ql\ and \ml, will be considerably strengthened at the LHC.

\begin {acknowledgments}

The research described here was initiated long, long ago when the authors were participating in the summer program at the Aspen Center for Physics (ACP).  They are grateful to the ACP for providing the scientific environments which stimulated this investigation.  The research of P.~L.~is supported by   an IBM Einstein Fellowship and by NSF grant PHYÐ0969448 and that of G.~S.~by DOE grant DE-FG02-91ER40690.  We thank John Beacom, James Beatty, Chris Hill, Andy Gould, Haibo Yu, and Grigory Rubtsov for enlightening discussions and communications.  We also thank the anonymous referee for suggesting that we explore the D0 results for an additional constraint on \ql.

\end{acknowledgments}
 
\appendix*

\section{Annihilation Cross Sections: Standard Model}

Here we consider annihilation into $f \bar f$ and $W^+W^-$ by $s$-channel $\gamma$ and $Z$ exchange,
into  $ZH$ by  $s$-channel $Z$ exchange, and into $\gamma\gamma, \gamma Z,$ and $ZZ$ by $t$- and $u$-channel $L$ exchange.  The results below and near the $Z$ pole use  the Breit-Wigner propagator $\left(s-M_Z^2+ i M_Z \Gamma_Z \right)^{-1}$.

\subsection{Annihilation to $f \bar f$}

Define $y_{f_L}$ ($y_{f_R}$) as the weak hypercharge of  the left (right) chiral projection of fermion $f$,  so that $q_f=t^3_{fL}+y_{f_L}= y_{f_R}$ ($y_{f_L}=1/6$ for the quarks and $-1/2$ for the leptons).  The $U(1)_Y$ boson couples with strength  $g' y_{f_L}$ to $f_L$, with strength $g' y_{f_R}$ to $f_R$, and with strength $g' y_L = g' Q_L$ to $L$.  The full expression, including $m_f$, $M_Z$, and $\Gamma_Z$ is too complicated to give here, but is included in the plots. In the limit $m_L \gg m_f, M_Z$, 
\beq   
(\sigma v)_{ ff}=\frac{C_f g^4 Q_L^2 \tan ^4\theta _W
\left(y_{f_L}^2+q_f^2\right)}{32 \pi  m_L^2},
 \eeql{e4}
where $g'= g \tan \theta _W $ and $C_f=3$ (quarks) or 1 (leptons) is the color factor.  Note that $e= g \sin \theta _W$.
Eq. \ref{e4} is equivalent to the QED expression
\beq   
(\sigma v)_{ ff}^{\rm QED}=\frac{C_f \pi \alpha^2 Q_L^2 q_f^2}{m_L^2},
 \eeql{e4a}
except that (in this limit) it is the $U(1)_Y$ boson $B$ that is actually exchanged.  That is, for $m_L$ much larger than the electroweak scale, one can think of $W^3$ and $B$ exchange rather than $\gamma$ and $Z$ exchange. However, the $W^3$ does not couple to $L$ since it is an $SU(2)$ singlet.  Including a final mass $m_f< m_L$, 
\begin{multline}
(\sigma v)_{ ff} = \frac{C_f g^4 \beta _f Q_L^2 \tan ^4\theta _W}{128 \pi  m_L^2} \\
\x  \bigl(6 q_f r_f^2 y_{f_L}+\left(4-r_f^2\right)  y_{f_L}^2+q_f^2 
\left(4-r_f^2\right)\bigr),
\label{e5}
\end{multline}
where $r_f=m_f/m_L$ and $\beta_f=\sqrt{1-r_f^2}$, generalizing Eq.~\ref{mtqed}.   

The corresponding values of $b_{ff}^{\rm SM}$ for three families are shown as a function of $m_L$ in Figure~\ref{fermions}, including the effects of $M_Z$ and $\Gamma_Z$.

\subsection{Annihilation to $\gamma\gamma,\ \gamma Z,$ and $ZZ$}

The cross sections for these channels are
\beq
\begin{split}
(\sigma v)_{ \gamma\gamma}& =\frac{g^4 Q_L^4 \sin ^4\theta _W}{16 \pi  m_L^2}\\
(\sigma v)_{ \gamma Z}& =  \frac{g^4 Q_L^4 \left(4-r_Z^2\right) \sin ^4\theta _W \tan ^2\theta _W\theta(1-r_Z/2)}{32\pi  m_L^2}\\
(\sigma v)_{ ZZ}& =  \frac{g^4 Q_L^4 \left(1-r_Z^2\right){}^{3/2} \sin ^4\theta _W \tan ^4\theta_W\ \theta(1-r_Z)}{4 \pi  m_L^2 \left(2-r_Z^2\right){}^2},
\end{split}
\eeql{e6}   
where $r_Z\equiv M_Z/m_L$.  For $m_L \gg M_Z$, the sum reduces to
\beq
(\sigma v)_{ BB} = \frac{g^4 Q_L^4 \tan ^4\theta _W}{16 \pi  M_L^2},
\eeql{e7}
which is just the $BB$ cross section since $g^4  \tan ^4\theta _W = g'^4$.

\subsection{Annihilation to $ZH$, where $H$ is the Higgs boson}

The cross section for the $ZH$ annihilation channel is
\begin{multline}
 (\sigma v)_{ ZH}=\frac{g^4 k_f Q_L^2 \tan^4\theta _W}{256 \pi  m_L^3 }\\
\x \frac{16 +40 r_Z^2-2 r_H^2 
\left(4+r_Z^2\right)+r_H^4+r_Z^4}{ 16-8 r_Z^2+r_Z^4+ r_Z^2\gamma_Z^2},
\label{e8}
\end{multline}
where $r_H\equiv M_H/m_L$,  $\gamma_Z\equiv 
\Gamma_Z/m_L$, and
\beq 
k_f=\frac{\sqrt{4 m_L^2-\left(M_Z-M_H\right){}^2} \sqrt{4 m_L^2-
\left(M_H+M_Z\right){}^2}}{4 m_L}.
\eeql{e9}
For $m_L\gg M_Z, M_H$ this reduces to 
\beq  (\sigma v)_{ ZH}=
\frac{g^4 Q_L^2 \tan ^4\theta _W}{256 \pi  m_L^2}.
\eeql{e10}

\subsection{Annihilation to $W^+ W^-$}

The cross section is 
\begin{multline}
(\sigma v)_{WW}=\frac{g^4 Q_L^2\beta_f \sin ^4\theta _W}{64 \pi  m_L^2 }\\
\x \frac{ \left(4+16 r_W^2-3 r_W^6-17 r_W^4
\right) r_Z^2 \left(r_Z^2+\gamma_Z^2\right)}{ r_W^4\left(16-8 r_Z^2+r_Z^4+ r_Z^2\gamma_Z^2\right)},
\label{e11}
\end{multline}
where $r_W\equiv M_W/m_L$,  $r_Z\equiv M_Z/m_L$,  and $\beta_f=\sqrt{1-r_W^2}$.  For $m_L\gg M_W, M_Z$,
\beq (\sigma v)_{WW}=
\frac{g^4 Q_L^2 \tan ^4\theta _W}{256 \pi  m_L^2}.
\eeql{e12}
It is at first surprising that this is nonzero. For large $m_L$ the $\gamma$ and $Z$ contributions cancel, leading to a factor  $M_Z^2/m_L^2\ra 0$ in the amplitude (i.e., one expects only the linear combination corresponding to the $B$ to survive, but the $B$ doesn't couple to $W^+W^-$).  However, the polarization vectors for the longitudinal $W^\pm$ each vary as $m_L/M_W$ in that limit, leading to a finite amplitude. The same result can be derived using the equivalence theorem~\cite{Lee:1977eg} at high $m_L$.

\subsection{Annihilation total for two bosons}

The contributions of the diboson channels to $a^{\rm SM}$ and $b^{\rm SM}$  are shown as a function of $m_L$ in Figure~\ref{bosonssm}, including the effects of $M_Z$ and $\Gamma_Z$. The Higgs mass is taken to be 120 GeV.

 \subsection{Drell-Yan production of FCHAMPs and gauginos}
 
The high-energy parton-level cross section for FCHAMP ($L^+ L^-$) and charged gaugino ($\tilde W^+ \tilde W^-$)  production are given by
 \beq
\begin{split}
\sigma _{LL} &=\frac{g^4 Q_L^2 \tan ^4\theta _W \left(y_{i_L}^2+q_i^2\right)\, \beta _f \left(3- \beta _f^2\right)}{144 \pi  s}\\
\sigma _{\tilde W \tilde W} &=\frac{g^4 \left(q_i-y_{i_L}\right)^2\, \beta _f \left(3-\beta _f^2\right)}{144 \pi  s},
\end{split} \eeql{fchampprod}
respectively,  where $q_i$ and $y_{i_L}$ are the electric charge and (left-chiral) weak hypercharge of the initial parton, and $\beta_f=\sqrt{1-4 m^2/s}$ is the final velocity of the FCHAMP or gaugino (of mass $m$). In Eq. \ref{fchampprod} we have neglected the parton mass and assumed $s\gg M_Z^2$, in which case the production procedes by $s$-channel  $B$ and $W_3$ exchange, respectively.  The ratio $\sigma _{LL} /\sigma _{\tilde W \tilde W} $ is therefore found to be $\sim 0.17 Q_L^2$ ($ 0.05 Q_L^2$) for $u\bar u$ ($d\bar d$). We have verified numerically that these ratios continue to hold to a good approximation when the full expressions involving $Z$ and $\gamma$ exchange are utilized. Since $u\bar u$  dominates at the Tevatron, we assume that the ratio of $p \bar p$ cross sections for FCHAMP and gaugino production is $0.15\, Q_L^2$ in the $Q_L-m_L$ range  relevant to the limits described in \S\ref{acc}.


\begin{thebibliography}{99}

\bibitem{khlopov}
M.~Yu.~Khlopov, JETP\ Lett.\ {\bf 33}, 162 (1981)

\bibitem{derujula}
 A.~DeRujula, S.~L.~Glashow and U.~Sarid, Nucl.\ Phys.\ B\ {\bf 333}, 173 (1990)

\bibitem{dimopoulos}
 S.~Dimopoulos, D.~Eichler, R.~Esmailzadeh and G.~D.~Starkman, Phys.\ Rev.\ D\ {\bf 41}, 2388 (1990)

\bibitem{gould}
 A.~Gould, B.~T.~Draine, R.~W.~Romani and S.~Nussinov, Phys.\ Lett.\ B\ {\bf 238}, 337 (1990)

\bibitem{davidson}
 S.~Davidson, B.~Campbell and D.~Bailey, Phys.\ Rev.\ D\ {\bf 43}, 2314 (1991)

\bibitem{wolfram}
 S.~Wolfram, Phys.\ Lett.\ B\ {\bf 82}, 65 (1979)

\bibitem{steigman1979}
 G.~Steigman, Ann.\ Rev.\ Nucl.\ Part.\ Sci.\ {\bf 29}, 313 (1979)

\bibitem{goldberg}
 H.~Goldberg, Phys.\ Rev.\ Lett.\ {\bf 48}, 1518 (1982)

 \bibitem{Perl:2009zz}
  M.~L.~Perl, E.~R.~Lee and D.~Loomba,
  Ann.\ Rev.\ Nucl.\ Part.\ Sci.\  {\bf 59}, 47 (2009).

\bibitem{Georgi:1974sy}
  H.~Georgi and S.~L.~Glashow,
  Phys.\ Rev.\ Lett.\  {\bf 32}, 438 (1974).

\bibitem{GellMann:1976pg}
  M.~Gell-Mann, P.~Ramond and R.~Slansky,
  Rev.\ Mod.\ Phys.\  {\bf 50}, 721 (1978).

\bibitem{Wen:1985qj}
  X.~G.~Wen and E.~Witten,
  Nucl.\ Phys.\  B {\bf 261}, 651 (1985).

\bibitem{Athanasiu:1988uj}
  G.~G.~Athanasiu, J.~J.~Atick, M.~Dine and W.~Fischler,
  Phys.\ Lett.\  B {\bf 214}, 55 (1988).
 
 \bibitem{Schellekens:1989qb}
  A.~N.~Schellekens,
  Phys.\ Lett.\  B {\bf 237}, 363 (1990).

\bibitem{Antoniadis:1989zy}
  I.~Antoniadis, J.~R.~Ellis, J.~S.~Hagelin and D.~V.~Nanopoulos,
  Phys.\ Lett.\  B {\bf 231}, 65 (1989).

\bibitem{Faraggi:1990af}
  A.~E.~Faraggi,
  Phys.\ Rev.\  D {\bf 46}, 3204 (1992).

\bibitem{Chaudhuri:1995ve}
  S.~Chaudhuri, G.~Hockney and J.~D.~Lykken,
  Nucl.\ Phys.\  B {\bf 469}, 357 (1996)
  [arXiv:hep-th/9510241].
  
\bibitem{Cleaver:1998gc}
  G.~Cleaver, M.~Cvetic, J.~R.~Espinosa, L.~L.~Everett, P.~Langacker and J.~Wang,
  Phys.\ Rev.\  D {\bf 59}, 055005 (1999)
  [arXiv:hep-ph/9807479].

\bibitem{Cvetic:2001nr}
  M.~Cvetic, G.~Shiu and A.~M.~Uranga,
  Nucl.\ Phys.\  B {\bf 615}, 3 (2001)
  [arXiv:hep-th/0107166].

\bibitem{Cvetic:2002qa}
  M.~Cvetic, P.~Langacker and G.~Shiu,
  Phys.\ Rev.\  D {\bf 66}, 066004 (2002)
  [arXiv:hep-ph/0205252].
  
  \bibitem{Cvetic:2011iq}
  M.~Cvetic, J.~Halverson, P.~Langacker,
    [arXiv:1108.5187 [hep-ph]].

\bibitem{Dienes:1995sq}
  K.~R.~Dienes, A.~E.~Faraggi and J.~March-Russell,
  Nucl.\ Phys.\  B {\bf 467}, 44 (1996)
  [arXiv:hep-th/9510223].

\bibitem{Lykken:1996kc}
  J.~D.~Lykken,
  Nucl.\ Phys.\ Proc.\ Suppl.\  {\bf 52A}, 271 (1997)
  [arXiv:hep-th/9607144].
  
\bibitem{Christodoulides:2011zs}
  K.~Christodoulides, A.~E.~Faraggi, J.~Rizos,
  [arXiv:1104.2264 [hep-ph]].  

\bibitem{Ellis:2004cj}
  J.~R.~Ellis, V.~E.~Mayes and D.~V.~Nanopoulos,
  Phys.\ Rev.\  D {\bf 70}, 075015 (2004)
  [arXiv:hep-ph/0403144].
  
\bibitem{Benakli:1998ut}
  K.~Benakli, J.~R.~Ellis and D.~V.~Nanopoulos,
  Phys.\ Rev.\  D {\bf 59}, 047301 (1999)
  [arXiv:hep-ph/9803333].

\bibitem{Chang:1996vw}
  S.~Chang, C.~Coriano and A.~E.~Faraggi,
  Nucl.\ Phys.\  B {\bf 477}, 65 (1996)
  [arXiv:hep-ph/9605325].
  
\bibitem{Coriano:2001mg}
  C.~Coriano, A.~E.~Faraggi and M.~Plumacher,
  Nucl.\ Phys.\  B {\bf 614}, 233 (2001)
  [arXiv:hep-ph/0107053].
  
\bibitem{Holdom:1985ag}
  B.~Holdom,
  Phys.\ Lett.\  B {\bf 166}, 196 (1986).
  
\bibitem{Foot:2004pa}
  R.~Foot,
  Int.\ J.\ Mod.\ Phys.\  D {\bf 13}, 2161 (2004)
  [arXiv:astro-ph/0407623].

\bibitem{Kors:2004dx}
  B.~Kors and P.~Nath,
  Phys.\ Lett.\  B {\bf 586}, 366 (2004)
  [arXiv:hep-ph/0402047].
 
\bibitem{Feldman:2007wj}
  D.~Feldman, Z.~Liu and P.~Nath,
  Phys.\ Rev.\  D {\bf 75}, 115001 (2007)
  [arXiv:hep-ph/0702123].

\bibitem{Prinz:1998ua}
  A.~A.~Prinz {\it et al.},
  Phys.\ Rev.\ Lett.\  {\bf 81}, 1175 (1998)
  [arXiv:hep-ex/9804008].
  
\bibitem{Davidson:2000hf}
  S.~Davidson, S.~Hannestad and G.~Raffelt,
  JHEP {\bf 0005}, 003 (2000)
  [arXiv:hep-ph/0001179].
  
  \bibitem{haibo}
  S.~D.~McDermott, H.~-B.~Yu, K.~M.~Zurek,
  Phys.\ Rev.\  {\bf D83}, 063509 (2011).
  [arXiv:1011.2907 [hep-ph]].
  
  \bibitem{Marinelli:1982dg}
  M.~Marinelli, G.~Morpurgo,
  Phys.\ Rept.\  {\bf 85}, 161-258 (1982). 

 \bibitem{smith}
  P.~F.~Smith,
  Ann.\ Rev.\ Nucl.\ Part.\ Sci.\  {\bf 39}, 73-111 (1989).
  
   \bibitem{Perl:2004qc}
  M.~L.~Perl, E.~R.~Lee, D.~Loomba,
  Mod.\ Phys.\ Lett.\  {\bf A19}, 2595-2610 (2004).
  
 \bibitem{Perl:2001xi}
  M.~L.~Perl, P.~C.~Kim, V.~Halyo, E.~R.~Lee, I.~T.~Lee, D.~Loomba and K.~S.~Lackner,
  Int.\ J.\ Mod.\ Phys.\  A {\bf 16}, 2137 (2001)
  [arXiv:hep-ex/0102033].
  
  \bibitem{Fairbairn:2006gg}
  M.~Fairbairn, A.~C.~Kraan, D.~A.~Milstead, T.~Sjostrand, P.~Z.~Skands and T.~Sloan,
  Phys.\ Rept.\  {\bf 438}, 1 (2007).
  [arXiv:hep-ph/0611040].
 
  \bibitem{abbiendi}
  G.~Abbiendi {\it et al.} [ OPAL Collaboration ], 
  Phys.\ Lett.\  {\bf B572}, 8-20 (2003).
  [hep-ex/0305031].
  
 \bibitem{Acosta:2002ju}
  D.~E.~Acosta {\it et al.} [ CDF Collaboration ],
  Phys.\ Rev.\ Lett.\  {\bf 90 } (2003)  131801.
  [hep-ex/0211064].
  
  \bibitem{Abazov:2008qu}
  V.~M.~Abazov {\it et al.} [ D0 Collaboration ],
  Phys.\ Rev.\ Lett.\  {\bf 102}, 161802 (2009).
  [arXiv:0809.4472 [hep-ex]].
  
  \bibitem{Aaltonen:2009kea}
  T.~Aaltonen {\it et al.} [ CDF Collaboration ],
  Phys.\ Rev.\ Lett.\  {\bf 103}, 021802 (2009).
  [arXiv:0902.1266 [hep-ex]].

\bibitem{Khachatryan:2011ts}
  V.~Khachatryan {\it et al.} [ CMS Collaboration ],
  JHEP {\bf 1103}, 024 (2011).
  [arXiv:1101.1645 [hep-ex]].
    
 \bibitem{Collaboration:2011hz}
The Atlas Collaboration,
    [arXiv:1106.4495 [hep-ex]].
 
\bibitem{komatsu}
 E.~Komatsu {\it et al.}, ApJS\ {\bf 192}, 18 (2011)

\bibitem{freedman}
 W.~Freedman {\it et al.}, ApJ\ {\bf 553}, 47 (2001)

\bibitem{mather}
 J.~Mather {\it et al.}, ApJ\ {\bf 512}, 511 (1999)

\bibitem{Gondolo:1990dk}
  P.~Gondolo, G.~Gelmini,
  Nucl.\ Phys.\  {\bf B360}, 145-179 (1991).

\bibitem{steigman2007}
 G.~Steigman, Ann.\ Rev.\ Nucl.\ Part.\ Sci.\ {\bf 57}, 463 (2007)
 
 \bibitem{rubtsov}
S.~L.~Dubovsky, D.~S.~Gorbunov and G.~I.~Rubtsov, JETP\ Lett. {\bf 79}, 1 (2004)

 \bibitem{nakamura}
K.~Nakamura {\it et al.} [Particle Data Group Collaboration],
J.\ Phys.\ G {\bf G37}, 075021 (2010)

 \bibitem{kim2007}
 P.~C.~Kim, E.~R.~Lee, I.~T.~Lee and M.~L.~Perl, 
 Phys.\ Rev.\ Lett.\ {\bf 99}, 161804 (2007).

\bibitem{lee}
I.~T.~Lee {\it et al.}, Phys.\ Rev.\ D {\bf 66}, 012002 (2002)

\bibitem{jones}
 W.~G.~Jones {\it et al.}, Z.\ Phys.\ C\ {\bf 43}, 349 (1989).

\bibitem{sommerfeld}
 A.~Sommerfeld, Ann.\ Phys.\ {\bf 403}, 257 (1931)

\bibitem{peebles}
 P.~J.~E.~Peebles, ApJS\ {\bf 9}, 185 (1964)
 
\bibitem{gamow}
G.~Gamow, Z.\ Phys.\ {\bf 51}, 204 (1928)

\bibitem{Lee:1977eg}
  B.~W.~Lee, C.~Quigg, H.~B.~Thacker,
  Phys.\ Rev.\  {\bf D16}, 1519 (1977)

\end{thebibliography}
\end{document}